\documentclass[aps,prl,reprint,nolongbibliography,noeprint,superscriptaddress]{revtex4-2}
\usepackage{amsfonts,amssymb,amsmath,bm,graphicx,color}
\usepackage[colorlinks=true,citecolor=blue,linkcolor=blue,urlcolor=blue]{hyperref}
\allowdisplaybreaks[4]
\DeclareMathOperator{\re}{Re}
\DeclareMathOperator{\im}{Im}
\DeclareMathOperator{\tr}{tr}
\newcommand{\db}[1]{\dot{\bm{#1}}}
\newcommand{\hb}[1]{\hat{\bm{#1}}}
\newcommand{\hc}[1]{\hat{\mathcal{#1}}}

\begin{document}
%--- Front Matter
\title{Spin accumulation in the spin Nernst effect}
\author{Atsuo Shitade}
\affiliation{Institute for Molecular Science, Aichi 444-8585, Japan}
\date{\today}
\begin{abstract}
  The spin Nernst effect is a phenomenon in which the spin current flows perpendicular to a temperature gradient.
  Similar to the spin Hall effect, this phenomenon also causes spin accumulation at the boundaries.
  Here, we study the spin response to the gradient of the temperature gradient with the use of Green's functions.
  Our formalism predicts physically observable spin accumulation without the ambiguity regarding the definition of the spin current or the magnetization correction.
  We prove the generalized Mott relation between the electric and thermal responses assuming the presence of time-reversal symmetry and the absence of inelastic scattering.
  We also find that thermal spin accumulation vanishes for the three-dimensional Luttinger model but is nonzero for the two-dimensional Rashba model
  with $\delta$-function nonmagnetic disorder in the first Born approximation.
\end{abstract}
\maketitle
\begin{figure}
  \centering
  \includegraphics[clip,width=0.48\textwidth]{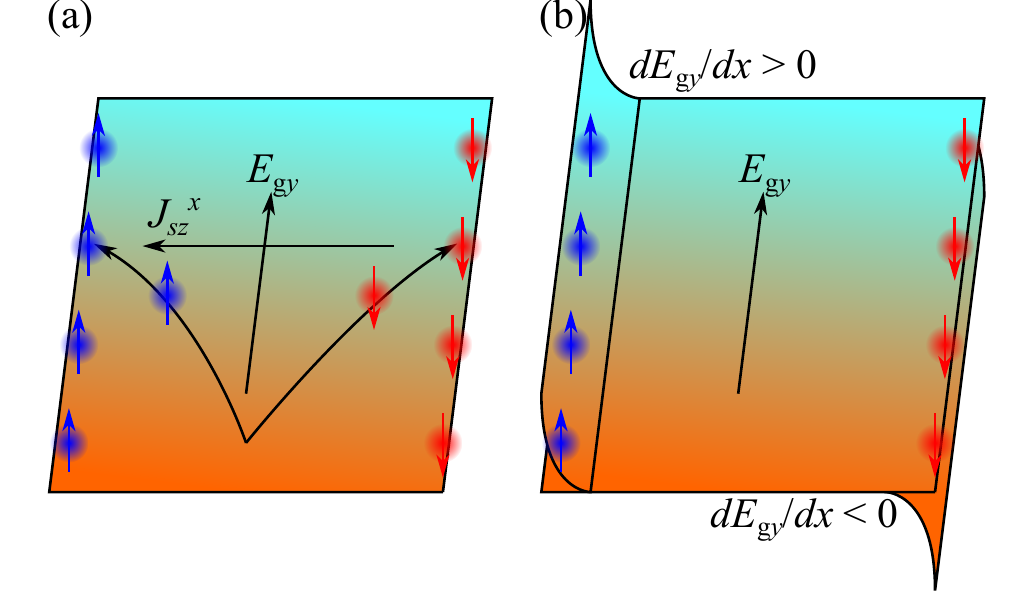}
  \caption{%
  (a) Typical scenario:
  The spin current is generated by a uniform temperature gradient via the SN effect and then turns into spin at the boundaries.
  (b) Our scenario:
  Spin is induced by the gradient of the temperature gradient at the boundaries.
  Our theory is free from the ambiguity regarding the definition of the spin current and spin torque or the difficulty of the magnetization correction.
  } \label{fig:sne}
\end{figure}

%--- Main Matter
\paragraph{Introduction.}
Spin caloritronics is a subfield of spintronics and aims to convert waste heat into spin that carries information~\cite{Bauer2012,JPSJ.90.122001}.
The door to this new research field was opened by the celebrated discovery of the spin Seebeck effect~\cite{Uchida2008,Jaworski2010},
in which the spin current is induced parallel to an applied temperature gradient.
On the other hand, the spin Nernst (SN) effect~\cite{BOSE2019165526} is a heat analog of the spin Hall (SH) effect~\cite{RevModPhys.87.1213},
namely, a phenomenon in which the spin current flows perpendicular to the temperature gradient and then turns into spin at the boundaries, as depicted in Fig.~\ref{fig:sne}(a).
This phenomenon was theoretically proposed~\cite{PhysRevB.78.045302,LIU2010471,MA2010510}
and experimentally observed in heavy metals such as platinum and tungsten using SN magnetoresistance~\cite{Meyer2017,Shenge1701503,ncomms1400}.
It will enrich candidate materials and device design for spincaloritronics.

The Landauer-B\"uttiker formalism is a powerful theoretical approach to the SH~\cite{PhysRevLett.94.016602,PhysRevB.72.075361} and SN effects~\cite{PhysRevB.78.045302,LIU2010471}.
In this formalism, the spin current is measured only in the leads, where a spin-orbit coupling is assumed to be absent, and hence is well defined.
The generalized Mott relation between the SH conductance~\cite{PhysRevLett.94.016602,PhysRevB.72.075361} and SN coefficient~\cite{PhysRevB.78.045302,LIU2010471} is obvious although not mentioned.

A complementary theory is based on the linear-response theory.
However, there are two difficulties in the field-theoretical formulation of the SN effect.
One is the ambiguity regarding the definition of the spin current.
In the presence of a spin-orbit coupling, spin is not conserved, and the spin current is not well defined.
We may use the conventional spin current, $\hat{J}_{sa}^{\phantom{sa} i}(\bm{k}) = \{\hat{s}_{a}, \hat{v}^{i}(\bm{k})\}/2$,
but this current is unphysical because its uniform equilibrium expectation value is nonzero in noncentrosymmetric systems such as the Rashba and Dresselhaus models~\cite{PhysRevB.68.241315}.
Here, $\hat{s}_{a}$ and $\hat{v}^{i}(\bm{k})$ are the spin and velocity operators, respectively.
We may also use the conserved spin current~\cite{PhysRevLett.96.076604,PhysRevB.77.075304}.
If the spin torque density $\tau_{a}(t, \bm{x})$ vanishes on average over the whole system,
we can define the spin torque dipole density as $\tau_{a}(t, \bm{x}) = -\partial_{x^{i}} P_{\tau a}^{\phantom{\tau a} i}(t, \bm{x})$,
and $\tilde{J}_{sa}^{\phantom{sa} i}(t, \bm{x}) = J_{sa}^{\phantom{sa} i}(t, \bm{x}) + P_{\tau a}^{\phantom{\tau a} i}(t, \bm{x})$ is conserved on average.
This current has a desirable property of Onsager's reciprocity relation.
However, there is no convincing way to define the spin current, since spin nonconservation is inevitable.

The other difficulty is magnetization correction.
According to one of the first theoretical proposals~\cite{MA2010510}, the Kubo formula of the SN conductivity of the conventional spin current diverges towards zero temperature.
Such unphysical divergence is well known in the context of the Nernst and thermal Hall effects
and can be corrected by orbital and heat magnetizations, respectively~\cite{0022-3719-10-12-021,PhysRevB.55.2344,PhysRevLett.106.197202,PhysRevB.84.184406,PhysRevLett.107.236601}.
These corrections arise because Luttinger's gravitational potential~\cite{PhysRev.135.A1505} perturbs the charge and heat current densities as well as the density matrix, leading to the Kubo formulas.
In the absence of inelastic scattering, the correct Nernst and thermal Hall conductivities are related to the Hall conductivity via the generalized Mott relation and Wiedemann-Franz law, respectively.
Similarly, the spin magnetic quadrupole moment is necessary for the gravitomagnetoelectric effect~\cite{PhysRevB.99.024404,PhysRevLett.124.066601},
in which the magnetization is induced by a temperature gradient.
Since the SN effect is a spin analog of the Nernst effect, that of orbital magnetization will be necessary for the SN effect.

So far, the SN conductivity of the conventional spin current has been computed assuming the generalized Mott relation
in first-principles calculations~\cite{PhysRevB.88.201108,PhysRevB.96.224415,PhysRevB.100.169907,PhysRevB.101.064430,Zhang_2020,PhysRevMaterials.4.124205},
but its justification remains unsettled.
In Refs.~\cite{PhysRevB.94.035306,PhysRevB.94.205302}, using the quantum-mechanical linear-response theory,
the authors introduced the spin-resolved orbital magnetization and obtained the SN conductivity that vanishes at zero temperature.
More strongly, it was recently shown that the nonequilibrium part satisfies the generalized Mott relation
using the semiclassical theory of wave packet dynamics~\cite{PhysRevLett.124.066601} and the non-Abelian Keldysh formalism~\cite{PhysRevB.105.035302}.
However, the equilibrium part does not always take the form of rotation, and hence its integral over a cross section does not vanish.
Regarding the conserved spin current, the equilibrium part takes the form of rotation, and the nonequilibrium part satisfies the generalized Mott relation~\cite{PhysRevB.98.081401,PhysRevB.104.L241411}.
These properties are desirable but not sufficient to define the spin current.

In the context of the SH effect, it is well understood that the essence is the spin accumulation, which is well defined theoretically and directly observable experimentally.
This idea was pointed out already in the first theoretical proposal~\cite{Dyakonov1971},
and the spin accumulation has been calculated via the spin diffusion equation and Landauer-Keldysh formalism~\cite{PhysRevB.70.195343,PhysRevLett.95.046601,PhysRevB.72.081301,PhysRevB.73.075303,RASHBA200631,PhysRevB.74.035340,PhysRevLett.99.106601,PhysRevLett.102.196802,PhysRevB.81.113304}.
However, it is difficult to deal with finite geometries for realistic multiband systems with complicated spin-orbit couplings.
Recently, a complementary theory was proposed considering the spin response to an electric-field gradient~\cite{PhysRevB.83.144428,PhysRevB.98.174422,PhysRevB.105.L201202}.
The electric field and resulting charge current must vanish outside the system and vary near the boundaries.
The gradient captures the lowest order of such a spatial variation and is considered to induce spin accumulation in this theory.
Since a nonuniform field can be described by the nonzero wave number under periodic boundary conditions,
this theory can be implemented to first-principles calculations.

In this Letter, we study the thermal spin accumulation at the boundaries in the SN effect by calculating the linear response of spin to a gradient of a temperature gradient, as depicted in Fig.~\ref{fig:sne}(b).
We prove the generalized Mott relation assuming the presence of time-reversal symmetry and the absence of inelastic scattering only.
The aforementioned difficulties can be avoided in this formulation;
we do not rely on the ambiguous spin current, and no correction term is necessary because the spin accumulation is a dissipative Fermi-surface term.
We apply this relation to the three-dimensional isotropic Luttinger model~\cite{PhysRev.102.1030} for $p$-type semiconductors and the two-dimensional Rashba model for $n$-type semiconductor heterostructures.
We find that thermal spin accumulation vanishes for the former but is nonzero for the latter.
Onsager's reciprocity relations are discussed.

In order to implement a temperature gradient $-\bm{\nabla}_{x} T$, which is a statistical force, into the linear-response theory,
we use a gravitational electric field $\bm{E}_{\mathrm{g}}$ originally introduced by Luttinger~\cite{PhysRev.135.A1505}.
This field is a mechanical force corresponding to the temperature gradient as $\bm{E}_{\mathrm{g}} \sim -\bm{\nabla}_{x} T/T$.
Similar to an electric field, this field is defined by either the gradient of Luttinger's gravitational potential or the time derivative of a gravitational vector potential~\cite{Shitade01122014,JPSJ.86.054601}.
The former is coupled to the Hamiltonian,
while the latter is coupled to the heat current and allows us to evaluate the spin--heat-current correlation function as the spin response to (the gradient of) the temperature gradient.

\paragraph{Generalized Mott relation.}
We calculate the linear response of $\hat{s}_{a}$ to the (gravitational) electric-field gradient,
\begin{equation}
  \langle \Delta \hat{s}_{a} \rangle^{(1, 1)}(\Omega, \bm{Q})
  = g_{a}^{\phantom{a} ij} (i Q_{i}) E_{j}(\Omega, \bm{Q}) + T g_{qa}^{\phantom{qa} ij} (i Q_{i}) E_{\mathrm{g} j}(\Omega, \bm{Q}). \label{eq:mott1}
\end{equation}
Since the (gravitational) electric field is defined as $\bm{E}_{\mathrm{(g)}}(\Omega, \bm{Q}) = -(-i \Omega) \bm{A}_{\mathrm{(g)}}(\Omega, \bm{Q})$
using the (gravitational) vector potential $\bm{A}_{\mathrm{(g)}}$,
we focus on the first-order terms with respect to the frequency $\Omega$ and wave number $\bm{Q}$ in the spin--charge- (heat-)current correlation function, which is implied by the superscript $(1, 1)$.
The temperature $T$ in the second term is introduced to compensate the denominator of $\bm{E}_{\mathrm{g}} \sim -\bm{\nabla}_{x} T/T$.
Thus, $g_{qa}^{\phantom{qa} ij}$ characterizes the thermal response $\langle \Delta \hat{s}_{a} \rangle^{(1, 1)} = g_{qa}^{\phantom{qa} ij} \partial_{x^{i}} (-\partial_{x^{j}} T)$.

In the presence of time-reversal symmetry, only the dissipative Fermi-surface term contributes to spin accumulation as
\begin{subequations} \begin{align}
  g_{a}^{\phantom{a} ij}
  = & \int d \epsilon
  [-f^{\prime}(\epsilon)] g_{0a}^{\phantom{0a} ij}(\epsilon), \label{eq:mott2a} \\
  T g_{qa}^{\phantom{qa} ij} 
  = & \int d \epsilon
  [-f^{\prime}(\epsilon)] T g_{q0a}^{\phantom{q0a} ij}(\epsilon), \label{eq:mott2b}
\end{align} \label{eq:mott2}\end{subequations}
in which $f(\epsilon) = [e^{(\epsilon - \mu)/T} + 1]^{-1}$ is the Fermi distribution function, and
\begin{widetext}
\begin{subequations} \begin{align}
  g_{0a}^{\phantom{0a} ij}(\epsilon)
  = & \frac{i \hbar^{2} q}{8 \pi}
  \int \frac{d^{d} k}{(2 \pi)^{d}}
  \tr [-2 \hat{S}_{a}^{\mathrm{AR}} \hat{G}^{\mathrm{R}}
  (\hat{V}^{\mathrm{RR} i} \hat{G}^{\mathrm{R}} \hat{V}^{\mathrm{RA} j} - \hat{V}^{\mathrm{RA} j} \hat{G}^{\mathrm{A}} \hat{V}^{\mathrm{AA} i}) \hat{G}^{\mathrm{A}} \notag \\
  & + \hat{S}_{a}^{\mathrm{AA}} \hat{G}^{\mathrm{A}}
  (\hat{V}^{\mathrm{AA} i} \hat{G}^{\mathrm{A}} \hat{V}^{\mathrm{AA} j} - \hat{V}^{\mathrm{AA} j} \hat{G}^{\mathrm{A}} \hat{V}^{\mathrm{AA} i}) \hat{G}^{\mathrm{A}}
  + \hat{S}_{a}^{\mathrm{RR}} \hat{G}^{\mathrm{R}}
  (\hat{V}^{\mathrm{RR} i} \hat{G}^{\mathrm{R}} \hat{V}^{\mathrm{RR} j} - \hat{V}^{\mathrm{RR} j} \hat{G}^{\mathrm{R}} \hat{V}^{\mathrm{RR} i}) \hat{G}^{\mathrm{R}}], \label{eq:mott3a} \\
  T g_{q0a}^{\phantom{q0a} ij}(\epsilon)
  = & \frac{i \hbar^{2}}{8 \pi}
  \int \frac{d^{d} k}{(2 \pi)^{d}}
  \tr [-2 \hat{S}_{a}^{\mathrm{AR}} \hat{G}^{\mathrm{R}}
  (\hat{V}^{\mathrm{RR} i} \hat{G}^{\mathrm{R}} \hc{J}_{q}^{\mathrm{RA} j} - \hc{J}_{q}^{\mathrm{RA} j} \hat{G}^{\mathrm{A}} \hat{V}^{\mathrm{AA} i}) \hat{G}^{\mathrm{A}} \notag \\
  & + \hat{S}_{a}^{\mathrm{AA}} \hat{G}^{\mathrm{A}}
  (\hat{V}^{\mathrm{AA} i} \hat{G}^{\mathrm{A}} \hc{J}_{q}^{\mathrm{AA} j} - \hc{J}_{q}^{\mathrm{AA} j} \hat{G}^{\mathrm{A}} \hat{V}^{\mathrm{AA} i}) \hat{G}^{\mathrm{A}}
  + \hat{S}_{a}^{\mathrm{RR}} \hat{G}^{\mathrm{R}}
  (\hat{V}^{\mathrm{RR} i} \hat{G}^{\mathrm{R}} \hc{J}_{q}^{\mathrm{RR} j} - \hc{J}_{q}^{\mathrm{RR} j} \hat{G}^{\mathrm{R}} \hat{V}^{\mathrm{RR} i}) \hat{G}^{\mathrm{R}}]. \label{eq:mott3b}
\end{align} \label{eq:mott3}\end{subequations}
\end{widetext}
We have omitted the arguments of $\epsilon$ and $\bm{k}$ for simplicity.
Equation~\eqref{eq:mott3a} was derived previously~\cite{PhysRevB.105.L201202}, while Eq.~\eqref{eq:mott3b} is a new result.
Here, $q$ is the electron charge, and $d$ is the spatial dimension.
$\hat{G}^{\alpha}(\epsilon, \bm{k}) = [\epsilon - \hat{\Sigma}^{\alpha}(\epsilon, \bm{k}) - \hc{H}(\bm{k})]^{-1}$ ($\alpha = \mathrm{R}, \mathrm{A}$) is the renormalized Green's function
with the self-energy $\hat{\Sigma}^{\alpha}(\epsilon, \bm{k})$.
$\hat{S}_{a}^{\alpha \beta}(\epsilon, \bm{k})$, $\hat{V}^{\alpha \beta i}(\epsilon, \bm{k})$, and $\hc{J}_{q}^{\alpha \beta j}(\epsilon, \bm{k})$
are the renormalized vertices of the spin, velocity, and heat current, respectively.
See Eqs.~\eqref{eq:conv_green3} and \eqref{eq:spin_green3} for a specific case.
The bare heat current vertex should be defined by the product of the time derivative and velocity~\cite{PhysRevB.67.014408,PhysRevB.94.104417}, rather than that of the Hamiltonian and velocity.
We do not specify either the Hamiltonian or the self-energy, and hence Eq.~\eqref{eq:mott3} can be applied to any disordered or interacting system as far as the time-reversal symmetry is preserved.

Hereafter, we restrict ourselves to disordered systems without inelastic scattering.
Since elastic scattering does not carry energy, the renormalized heat current vertex $\hc{J}_{q}^{\alpha \beta j}(\epsilon, \bm{k})$ is reduced to
$\hc{J}_{q}^{\alpha \beta j}(\epsilon, \bm{k}) = (\epsilon - \mu) \hat{V}^{\alpha \beta j}(\epsilon, \bm{k})$.
Then, we obtain $T g_{q0a}^{\phantom{q0a} ij}(\epsilon) = (\epsilon - \mu) g_{0a}^{\phantom{0a} ij}(\epsilon)/q$,
and Eq.~\eqref{eq:mott2} becomes the generalized Mott relation, which is reduced to the Mott relation near zero temperature using the Sommerfeld expansion.
This is our main result.
The only assumptions are the presence of time-reversal symmetry and the absence of inelastic scattering.

\paragraph{Luttinger model.}
First, we consider the three-dimensional isotropic Luttinger model~\cite{PhysRev.102.1030} for $p$-type semiconductors,
which was studied in the context of the SH effect~\cite{Murakami1348,PhysRevB.69.235206}.
The Hamiltonian is expressed as
\begin{equation}
  \hc{H}(\bm{k})
  = \frac{\hbar^{2}}{2 m} \left[\left(\gamma_{1} + \frac{5}{2} \gamma_{2}\right) k^{2} - 2 \gamma_{2} (\bm{k} \cdot \bm{\Sigma})^{2}\right], \label{eq:luttinger1}
\end{equation}
in which $\bm{\Sigma}$ is the $j = 3/2$ matrix.
We can construct $\Gamma$ matrices satisfying $\{\Gamma_{a}, \Gamma_{b}\} = 2 \delta_{ab}$ ($a = 1, \dots, 5$) from products of $\bm{\Sigma}$~\cite{PhysRevB.69.235206}.
Using these $\Gamma$ matrices, this Hamiltonian can be rewritten as
\begin{equation}
  \hc{H}(\bm{k})
  = \frac{\hbar^{2} k^{2}}{2 m} \left[\gamma_{1} + 2 \gamma_{2} \sum_{a = 1}^{5} h^{a}(\bm{k}) \Gamma_{a}\right], \label{eq:luttinger4}
\end{equation}
in which
\begin{subequations} \begin{align}
  h^{1}(\bm{k})
  = & -\sqrt{3} \sin \theta \cos \theta \sin \phi, \label{eq:luttinger5a} \\
%%%%%%%%%%
  h^{2}(\bm{k})
  = & -\sqrt{3} \sin \theta \cos \theta \cos \phi, \label{eq:luttinger5b} \\
%%%%%%%%%%
  h^{3}(\bm{k})
  = & -\frac{\sqrt{3}}{2} \sin^{2} \theta \sin 2 \phi, \label{eq:luttinger5c} \\
%%%%%%%%%%
  h^{4}(\bm{k})
  = & -\frac{\sqrt{3}}{2} \sin^{2} \theta \cos 2 \phi, \label{eq:luttinger5d} \\
%%%%%%%%%%
  h^{5}(\bm{k})
  = & -\frac{1}{2} (3 \cos^{2} \theta - 1). \label{eq:luttinger5e}
\end{align} \label{eq:luttinger5}\end{subequations}
The eigenvalues are $\epsilon_{\sigma}(\bm{k}) = (\gamma_{1} + 2 \sigma \gamma_{2}) \hbar^{2} k^{2}/2 m$ with $\sigma = \pm 1$,
each of which is doubly degenerate owing to the time-reversal and inversion symmetries.
$\sigma = +1$ is called a light hole, while $\sigma = -1$ is called a heavy hole.

We consider $\delta$-function nonmagnetic disorder within the first Born approximation.
The bare retarded Green's function is expressed as
\begin{align}
  \hat{g}^{\mathrm{R}}(\epsilon, \bm{k})
  = & \frac{1}{\epsilon + i \eta - \hc{H}(\bm{k})} \notag \\
  = & \frac{1}{2} [g_{+}^{\mathrm{R}}(\epsilon, \bm{k}) + g_{-}^{\mathrm{R}}(\epsilon, \bm{k})] \notag \\
  & + \frac{1}{2} [g_{+}^{\mathrm{R}}(\epsilon, \bm{k}) - g_{-}^{\mathrm{R}}(\epsilon, \bm{k})] \sum_{a = 1}^{5} h^{a}(\bm{k}) \Gamma_{a}, \label{eq:green1}
\end{align}
with $g_{\sigma}^{\mathrm{R}}(\epsilon, \bm{k}) = [\epsilon + i \eta - \epsilon_{\sigma}(\bm{k})]^{-1}$.
In the first Born approximation, the imaginary part of the self-energy is
\begin{equation}
  \hat{\Gamma}(\epsilon)
  = -\im \left[n_{\mathrm{i}} v_{\mathrm{i}}^{2}
  \int \frac{d^{3} k}{(2 \pi)^{3}}
  \hat{g}^{\mathrm{R}}(\epsilon, \bm{k})\right]
  = \pi n_{\mathrm{i}} v_{\mathrm{i}}^{2} D(\epsilon), \label{eq:green2}
\end{equation}
in which
\begin{align}
  D(\epsilon)
  = & \frac{1}{2}
  \sum_{\sigma} \int_{0}^{\infty} \frac{k^{2} d k}{2 \pi^{2}}
  \frac{-1}{\pi} \im [g_{\sigma}^{\mathrm{R}}(\epsilon, \bm{k})] \notag \\
  = & \frac{m}{4 \pi^{2} \hbar^{2}}
  \sum_{\sigma}
  \frac{1}{(\gamma_{1} + 2 \sigma \gamma_{2})^{3/2}} \sqrt{\frac{2 m \epsilon}{\hbar^{2}}}, \label{eq:green3}
\end{align}
is the density of states.
Below, we denote $\hat{\Gamma}(\epsilon)$ as $\Gamma(\epsilon)$.
The renormalized retarded Green's function is given as
\begin{align}
  \hat{G}^{\mathrm{R}}(\epsilon, \bm{k})
  = & \frac{1}{\epsilon + i \Gamma(\epsilon) - \hc{H}(\bm{k})} \notag \\
  = & \frac{1}{2} [G_{+}^{\mathrm{R}}(\epsilon, \bm{k}) + G_{-}^{\mathrm{R}}(\epsilon, \bm{k})] \notag \\
  & + \frac{1}{2} [G_{+}^{\mathrm{R}}(\epsilon, \bm{k}) - G_{-}^{\mathrm{R}}(\epsilon, \bm{k})] \sum_{a = 1}^{5} h^{a}(\bm{k}) \Gamma_{a}, \label{eq:green4}
\end{align}
with $G_{\sigma}^{\mathrm{R}}(\epsilon, \bm{k}) = [\epsilon + i \Gamma(\epsilon) - \epsilon_{\sigma}(\bm{k})]^{-1}$.

We take into account the vertex corrections.
First, $\hat{V}^{\mathrm{RR} i}(\epsilon, \bm{k}) = \hat{V}^{\mathrm{AA} i}(\epsilon, \bm{k}) = \hat{v}^{i}(\bm{k}) = \hbar^{-1} \partial_{k_{i}} \hc{H}(\bm{k})$,
since $\delta$-function disorder does not carry a wave number.
Second, $\hat{V}^{\mathrm{RA} y}(\epsilon, \bm{k})$ is obtained by solving
\begin{align}
  \hat{V}^{\mathrm{RA} y}(\epsilon, \bm{k})
  = & \hat{v}^{y}(\bm{k}) + n_{\mathrm{i}} v_{\mathrm{i}}^{2}
  \int \frac{d^{3} k^{\prime}}{(2 \pi)^{3}} \notag \\
  & \times \hat{G}^{\mathrm{R}}(\epsilon, \bm{k}^{\prime}) \hat{V}^{\mathrm{RA} y}(\epsilon, \bm{k}^{\prime}) \hat{G}^{\mathrm{A}}(\epsilon, \bm{k}^{\prime}), \label{eq:conv_green3}
\end{align}
but results in $\hat{V}^{\mathrm{RA} y}(\epsilon, \bm{k}) = \hat{v}^{y}(\bm{k})$ owing to the inversion symmetry~\cite{PhysRevB.69.241202}.
$\hat{S}_{z}^{\mathrm{AR}}(\epsilon, \bm{k})$ is obtained by solving
\begin{align}
  \hat{S}_{z}^{\mathrm{AR}}(\epsilon, \bm{k})
  = & \hat{s}_{z} + n_{\mathrm{i}} v_{\mathrm{i}}^{2}
  \int \frac{d^{3} k^{\prime}}{(2 \pi)^{3}} \notag \\
  & \times \hat{G}^{\mathrm{A}}(\epsilon, \bm{k}^{\prime}) \hat{S}_{z}^{\mathrm{AR}}(\epsilon, \bm{k}^{\prime}) \hat{G}^{\mathrm{R}}(\epsilon, \bm{k}^{\prime}). \label{eq:spin_green3}
\end{align}
We put $\hat{s}_{z} = (\hbar/3) \Sigma_{z}$, because the contribution of the spin $s = 1/2$ to the total angular momentum $j = 3/2$ is $1/3$~\cite{Murakami1348}.
The solution takes the form of $\hat{S}_{z}^{\mathrm{AR}}(\epsilon, \bm{k}) = (\hbar/3) S_{z}^{z}(\epsilon) \Sigma_{z}$.
Thus, $g_{0z}^{\phantom{0z} xy}(\epsilon)$ is obtained as
\begin{widetext}
\begin{align}
  g_{0z}^{\phantom{0z} xy}(\epsilon)
  = & \frac{\gamma_{2} \hbar^{5} q}{12 \pi m^{2}}
  \int_{0}^{\infty} \frac{k^{2} d k}{2 \pi^{2}}
  \left(S_{z}^{z}(\epsilon) \{(\gamma_{1} + 2 \gamma_{2}) [(G_{+}^{\mathrm{A}})^{2} G_{-}^{\mathrm{R}} + G_{-}^{\mathrm{A}} (G_{+}^{\mathrm{R}})^{2}]
  - (\gamma_{1} - 2 \gamma_{2}) [(G_{-}^{\mathrm{A}})^{2} G_{+}^{\mathrm{R}} + G_{+}^{\mathrm{A}} (G_{-}^{\mathrm{R}})^{2}]\right. \notag \\
  & + (\gamma_{1} + \gamma_{2}) G_{+}^{\mathrm{A}} G_{+}^{\mathrm{R}} (G_{-}^{\mathrm{A}} + G_{-}^{\mathrm{A}})
  - (\gamma_{1} - 5 \gamma_{2}) G_{-}^{\mathrm{A}} G_{-}^{\mathrm{R}} (G_{+}^{\mathrm{A}} + G_{+}^{\mathrm{A}})\} \notag \\
  & \left.- (2 \gamma_{1} + 3 \gamma_{2}) [(G_{+}^{\mathrm{A}})^{2} G_{-}^{\mathrm{A}} + (G_{+}^{\mathrm{R}})^{2} G_{-}^{\mathrm{R}}]
  + (2 \gamma_{1} - 7 \gamma_{2}) [G_{+}^{\mathrm{A}} (G_{-}^{\mathrm{A}})^{2} + G_{+}^{\mathrm{R}} (G_{-}^{\mathrm{R}})^{2}]\right)k^{2} \notag \\
  = & \frac{\gamma_{2} \hbar^{3} q}{12 m}
  \frac{S_{z}^{z}(\epsilon)}{\Gamma(\epsilon)} \sum_{\sigma} \int_{0}^{\infty} \frac{k^{2} d k}{2 \pi^{2}}
  \frac{-1}{\pi} \im \left(\left\{3 \sigma - \frac{2 \gamma_{2} \hbar^{2} k^{2}/m}{\gamma_{2} \hbar^{2} k^{2}/m - i \sigma \Gamma(\epsilon)}
  + \frac{\gamma_{1} \gamma_{2} (\hbar^{2} k^{2}/m)^{2}}{[\gamma_{2} \hbar^{2} k^{2}/m - i \sigma \Gamma(\epsilon)]^{2}}\right\} G_{\sigma}^{\mathrm{R}}\right) \notag \\
  & + \frac{\hbar^{3} q}{12 m}
  \sum_{\sigma} \int_{0}^{\infty} \frac{k^{2} d k}{2 \pi^{2}}
  \frac{1}{\pi} \re \left\{\frac{\gamma_{2} S_{z}^{z}(\epsilon) k \partial_{k} G_{\sigma}^{\mathrm{R}}}{\gamma_{2} \hbar^{2} k^{2}/m - i \sigma \Gamma(\epsilon)}
  - [2 \gamma_{1} + (5 \sigma - 2) \gamma_{2}] (G_{\sigma}^{\mathrm{R}})^{2} + \frac{2 \sigma (\gamma_{1} - \gamma_{2}) G_{\sigma}^{\mathrm{R}}}{\gamma_{2} \hbar^{2} k^{2}/m}\right\}, \label{eq:spin_green2b}
\end{align}
with
\begin{align}
  S_{z}^{z}(\epsilon)
  = & \left[1 - \frac{n_{\mathrm{i}} v_{\mathrm{i}}^{2}}{10}
  \int_{0}^{\infty} \frac{k^{2} d k}{2 \pi^{2}}
  (3 G_{+}^{\mathrm{A}} G_{+}^{\mathrm{R}} + 3 G_{-}^{\mathrm{A}} G_{-}^{\mathrm{R}} + 2 G_{+}^{\mathrm{A}} G_{-}^{\mathrm{R}} + 2 G_{-}^{\mathrm{A}} G_{+}^{\mathrm{R}})\right]^{-1} \notag \\
  = & \left(1 - \pi n_{\mathrm{i}} v_{\mathrm{i}}^{2}
  \sum_{\sigma} \int_{0}^{\infty} \frac{k^{2} d k}{2 \pi^{2}}
  \left\{\frac{3}{10 \Gamma(\epsilon)} \frac{-1}{\pi} \im (G_{\sigma}^{\mathrm{R}})
  + \frac{1}{5} \frac{1}{\pi} \re \left[\frac{\sigma G_{\sigma}^{\mathrm{R}}}{\gamma_{2} \hbar^{2} k^{2}/m - i \sigma \Gamma(\epsilon)}\right]\right\}\right)^{-1}. \label{eq:spin_green4}
\end{align}
\end{widetext}
We have omitted the arguments of $\epsilon$ and $\bm{k}$ for simplicity.

In the limit of $n_{\mathrm{i}} v_{\mathrm{i}}^{2} \rightarrow +0$, we find
\begin{equation}
  g_{0z}^{\phantom{0z} xy}(\epsilon)
  = \frac{q}{12 \pi^{2}} \frac{\hbar S_{z}^{z}(\epsilon)}{2 \Gamma(\epsilon)}
  \sum_{\sigma} \frac{\gamma_{1} + (3 \sigma - 2) \gamma_{2}}{(\gamma_{1} + 2 \sigma \gamma_{2})^{3/2}} \sqrt{\frac{2 m \epsilon}{\hbar^{2}}}, \label{eq:spin_green5b}
\end{equation}
with $S_{z}^{z}(\epsilon) = 5/2$.
Since the density of states~\eqref{eq:green3} is also proportional to $\sqrt{2 m \epsilon/\hbar^{2}}$, $g_{0z}^{\phantom{0z} xy}(\epsilon)$ is constant.
Hence, the electrical spin accumulation $g_{z}^{\phantom{z} xy}$ does not depend on $\mu$ or $T$,
and the thermal one $g_{qz}^{\phantom{qz} xy}$ vanishes.
Note that this vanishing response results from the accidental cancellation of the energy dependence.
The thermal response can be nonzero for long-range nonmagnetic or spin-orbit-coupled disorder or with higher-order approximations.

\begin{figure*}
  \centering
  \includegraphics[clip,width=0.98\textwidth]{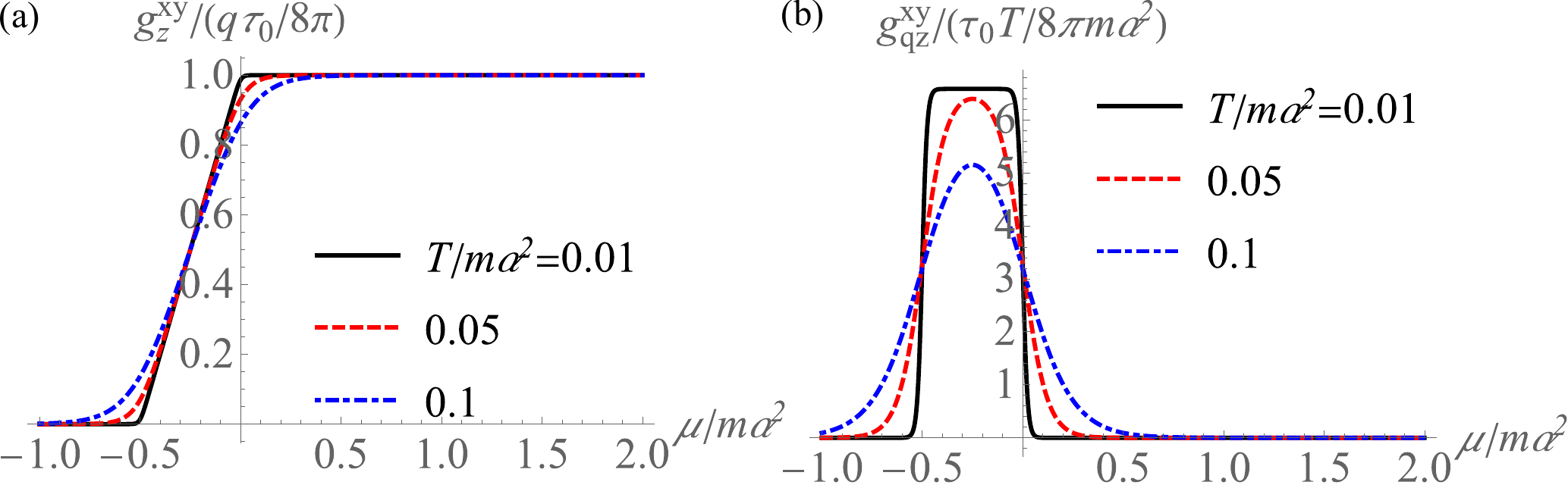}
  \caption{%
  The chemical potential dependences of spin accumulation induced
  (a) by an electric field $g_{z}^{\phantom{z} xy}$
  and (b) by a temperature gradient $g_{qz}^{\phantom{qz} xy}$.
  The black solid, red dashed, and blue dotted-dashed lines represent $T/m \alpha^{2} = 0.01$, $0.05$, and $0.1$, respectively.
  } \label{fig:rashba}
\end{figure*}

\paragraph{Rashba model.}
Second, we consider the two-dimensional Rashba model for $n$-type semiconductor heterostructures, which was also studied in the context of the SH effect~\cite{PhysRevLett.92.126603}.
The Hamiltonian is expressed as
\begin{equation}
  \hc{H}(\bm{k})
  = \frac{\hbar^{2} k^{2}}{2 m} + \hbar \alpha (k_{y} \sigma_{x} - k_{x} \sigma_{y}), \label{eq:rashba1}
\end{equation}
where $\bm{\sigma}$ is the Pauli matrix corresponding to the spin operator $\hb{s} = (\hbar/2) \bm{\sigma}$.
In the case of $\delta$-function nonmagnetic disorder within the first Born approximation, $g_{0z}^{\phantom{0z} xy}(\epsilon)$ was already obtained as~\cite{PhysRevB.105.L201202}
\begin{equation}
  g_{0z}^{\phantom{0z} xy}(\epsilon)
  = \frac{q \tau_{0}}{8 \pi}
  \begin{cases}
    1 & (\epsilon > 0), \\
    1 + 2 \epsilon/m \alpha^{2} & (\epsilon < 0).
  \end{cases} \label{eq:rashba2}
\end{equation}
Here, $\tau_{0} = \hbar/2 \Gamma_{0}$ is the relaxation time with $\Gamma_{0} = m n_{\mathrm{i}} v_{\mathrm{i}}^{2}/2 \hbar^{2}$.
The chemical potential dependences of $g_{(q)z}^{\phantom{(q)z} xy}$ are shown in Fig.~\ref{fig:rashba}.
Apart from the thermal excitation effect, $g_{qz}^{\phantom{qz} xy}$ is nonzero only below the Rashba crossing, where $g_{0z}^{\phantom{0z} xy}(\epsilon)$ depends on $\epsilon$.
The plateau at the lowest temperature in Fig.~\ref{fig:rashba}(b) takes the value of $2 \pi^{2}/3$ according to the Mott relation.

\paragraph{Discussion.}
It has been believed that the spin current can be measured via the dampinglike (DL) spin torque
using the ferromagnetic resonance in a ferromagnet--heavy-metal heterostructure~\cite{PhysRevLett.101.036601,PhysRevLett.106.036601}.
The unit vector of the magnetization $\bm{n}$ obeys the Landau-Lifshitz-Gilbert equation,
\begin{equation}
  \db{n}
  = \gamma \bm{H}_{\mathrm{eff}} \times \bm{n} + \alpha_{\mathrm{G}} \bm{n} \times \db{n} + \bm{\tau}^{\prime}, \label{eq:discussion1}
\end{equation}
in which $\gamma \bm{H}_{\mathrm{eff}}$ is the effective magnetic field including the gyromagnetic ratio, $\alpha_{\mathrm{G}}$ is the Gilbert damping,
and $\bm{\tau}^{\prime}$ is an additional spin torque.
The spin current is assumed to yield the DL spin transfer torque $\bm{\tau}^{\prime} \propto J_{s} \bm{n} \times (\bm{\sigma} \times \bm{n})$
with the direction of the spin polarization of the spin current $\bm{\sigma}$.
On the other hand, spin accumulation yields the fieldlike spin torque $\bm{\tau}^{\prime} = \bm{n} \times J \langle \Delta \hb{s} \rangle/\hbar$
if we consider the exchange coupling $\hc{H}^{\prime} = -J \bm{n} \cdot \hb{s}/\hbar$ near the interface.
The DL spin torque may arise in combination with the Gilbert damping.
In addition, since both the inversion and time-reversal symmetries are broken near the interface,
the magnetoelectric effect is allowed and contributes to the DL spin torque, as already pointed out in Ref.~\cite{RevModPhys.87.1213}.
Thus, it is difficult to extract the spin current from the DL spin torque.

Onsager's reciprocity relations readily follow from Eq.~\eqref{eq:mott3}.
Since spin is conjugate to the Zeeman field, namely, $\hc{H}^{\prime} = -\bm{B} \cdot \hb{s}$, the inverse SH and SN effects are described as
\begin{subequations} \begin{align}
  \langle \Delta \hat{J}^{j} \rangle^{(1, 1)}(\Omega, \bm{Q})
  = & g_{a}^{\phantom{a} ij} (i Q_{i}) \dot{B}^{a}(\Omega, \bm{Q}), \label{eq:discussion2a} \\
  \langle \Delta \hat{J}_{q}^{j} \rangle^{(1, 1)}(\Omega, \bm{Q})
  = & T g_{qa}^{\phantom{qa} ij} (i Q_{i}) \dot{B}^{a}(\Omega, \bm{Q}). \label{eq:discussion2b}
\end{align} \label{eq:discussion2}\end{subequations}
Equation~\eqref{eq:discussion2a} is similar to the previous formulation of the inverse Edelstein effect~\cite{PhysRevLett.112.096601}.
The external force $\db{B}$ can be realized by the spin pumping in a ferromagnet--heavy-metal heterostructure as follows:
We can read $\bm{B} = J \bm{n}/\hbar$ from the aforementioned exchange coupling near the interface.
Assuming a simple precession along the $a$ axis, there is a dc component in $\dot{B}^{a} = J \alpha_{\mathrm{G}} (\bm{n} \times \db{n})^{a}/\hbar$,
which is interpreted as the spin current in the spin pumping theory~\cite{PhysRevLett.88.117601,PhysRevB.66.224403,RevModPhys.77.1375}.
A dc component of $\db{B}$ also arises from the second-order perturbation with respect to the exchange coupling.
Indeed, the previous microscopic calculations revealed that the charge current is induced by
$\bm{n} \times \db{n}$~\cite{PhysRevLett.99.266603,JPSJ.77.074701,PhysRevB.81.144405} and $\partial_{x^{i}} (\bm{n} \times \db{n})$~\cite{PhysRevB.81.144405,JPSJ.79.014708},
which can be attributed to the inverse Edelstein and SH effects, respectively.

Finally, let us comment on the importance of time-reversal symmetry.
In the presence of this symmetry, as assumed above, the only process to exhibit spin accumulation is a dissipative Fermi-surface term.
That is why we do not suffer from a magnetization correction, which is a thermodynamic quantity.
If a system lacks time-reversal symmetry, a nondissipative Fermi-sea term is allowed.
Note that the SH effect in such a system is called the magnetic SH effect~\cite{Kimata2019}.
As in the Nernst and thermal Hall effects~\cite{0022-3719-10-12-021,PhysRevB.55.2344,PhysRevLett.106.197202,PhysRevB.84.184406,PhysRevLett.107.236601}
and the gravitomagnetoelectric effect~\cite{PhysRevB.99.024404,PhysRevLett.124.066601}, the Kubo formula of thermal spin accumulation diverges towards zero temperature.
Such unphysical divergence would be corrected by the spin magnetic octupole moment, which is a future problem.

\paragraph{Summary.}
To summarize, we have studied the spin response to the gradient of a temperature gradient
and proved the generalized Mott relation to the electric counterpart assuming the presence of time-reversal symmetry and the absence of inelastic scattering only.
Since such a gradient appears at the boundaries, this response describes the thermal spin accumulation.
In sharp contrast to previous theories on the SN effect, our theory is free from the ambiguity regarding the definition of the spin current or the magnetization correction.
We have found that thermal spin accumulation is absent in the isotropic Luttinger model for $p$-type semiconductors but present in the Rashba model for $n$-type semiconductor heterostructures.
Onsager's reciprocity relations are evident in our theory.

%--- Back Matter
\begin{acknowledgments}
  \paragraph{Acknowledgments.}
  This work was supported by the Japan Society for the Promotion of Science KAKENHI (Grants No.~JP21H01816 and No.~JP22K03498).
\end{acknowledgments}

\begin{thebibliography}{69}%
\makeatletter
\providecommand \@ifxundefined [1]{%
 \@ifx{#1\undefined}
}%
\providecommand \@ifnum [1]{%
 \ifnum #1\expandafter \@firstoftwo
 \else \expandafter \@secondoftwo
 \fi
}%
\providecommand \@ifx [1]{%
 \ifx #1\expandafter \@firstoftwo
 \else \expandafter \@secondoftwo
 \fi
}%
\providecommand \natexlab [1]{#1}%
\providecommand \enquote  [1]{``#1''}%
\providecommand \bibnamefont  [1]{#1}%
\providecommand \bibfnamefont [1]{#1}%
\providecommand \citenamefont [1]{#1}%
\providecommand \href@noop [0]{\@secondoftwo}%
\providecommand \href [0]{\begingroup \@sanitize@url \@href}%
\providecommand \@href[1]{\@@startlink{#1}\@@href}%
\providecommand \@@href[1]{\endgroup#1\@@endlink}%
\providecommand \@sanitize@url [0]{\catcode `\\12\catcode `\$12\catcode
  `\&12\catcode `\#12\catcode `\^12\catcode `\_12\catcode `\%12\relax}%
\providecommand \@@startlink[1]{}%
\providecommand \@@endlink[0]{}%
\providecommand \url  [0]{\begingroup\@sanitize@url \@url }%
\providecommand \@url [1]{\endgroup\@href {#1}{\urlprefix }}%
\providecommand \urlprefix  [0]{URL }%
\providecommand \Eprint [0]{\href }%
\providecommand \doibase [0]{https://doi.org/}%
\providecommand \selectlanguage [0]{\@gobble}%
\providecommand \bibinfo  [0]{\@secondoftwo}%
\providecommand \bibfield  [0]{\@secondoftwo}%
\providecommand \translation [1]{[#1]}%
\providecommand \BibitemOpen [0]{}%
\providecommand \bibitemStop [0]{}%
\providecommand \bibitemNoStop [0]{.\EOS\space}%
\providecommand \EOS [0]{\spacefactor3000\relax}%
\providecommand \BibitemShut  [1]{\csname bibitem#1\endcsname}%
\let\auto@bib@innerbib\@empty
%</preamble>
\bibitem [{\citenamefont {Bauer}\ \emph {et~al.}(2012)\citenamefont {Bauer},
  \citenamefont {Saitoh},\ and\ \citenamefont {van Wees}}]{Bauer2012}%
  \BibitemOpen
  \bibfield  {author} {\bibinfo {author} {\bibfnamefont {G.~E.~W.}\
  \bibnamefont {Bauer}}, \bibinfo {author} {\bibfnamefont {E.}~\bibnamefont
  {Saitoh}},\ and\ \bibinfo {author} {\bibfnamefont {B.~J.}\ \bibnamefont {van
  Wees}},\ }\href {https://doi.org/10.1038/nmat3301} {\bibfield  {journal}
  {\bibinfo  {journal} {Nat. Mater.}\ }\textbf {\bibinfo {volume} {11}},\
  \bibinfo {pages} {391} (\bibinfo {year} {2012})}\BibitemShut {NoStop}%
\bibitem [{\citenamefont {Uchida}\ and\ \citenamefont
  {Iguchi}(2021)}]{JPSJ.90.122001}%
  \BibitemOpen
  \bibfield  {author} {\bibinfo {author} {\bibfnamefont {K.-i.}\ \bibnamefont
  {Uchida}}\ and\ \bibinfo {author} {\bibfnamefont {R.}~\bibnamefont
  {Iguchi}},\ }\href {https://doi.org/10.7566/JPSJ.90.122001} {\bibfield
  {journal} {\bibinfo  {journal} {J. Phys. Soc. Jpn.}\ }\textbf {\bibinfo
  {volume} {90}},\ \bibinfo {pages} {122001} (\bibinfo {year}
  {2021})}\BibitemShut {NoStop}%
\bibitem [{\citenamefont {Uchida}\ \emph {et~al.}(2008)\citenamefont {Uchida},
  \citenamefont {Takahashi}, \citenamefont {Harii}, \citenamefont {Ieda},
  \citenamefont {Koshibae}, \citenamefont {Ando}, \citenamefont {Maekawa},\
  and\ \citenamefont {Saitoh}}]{Uchida2008}%
  \BibitemOpen
  \bibfield  {author} {\bibinfo {author} {\bibfnamefont {K.}~\bibnamefont
  {Uchida}}, \bibinfo {author} {\bibfnamefont {S.}~\bibnamefont {Takahashi}},
  \bibinfo {author} {\bibfnamefont {K.}~\bibnamefont {Harii}}, \bibinfo
  {author} {\bibfnamefont {J.}~\bibnamefont {Ieda}}, \bibinfo {author}
  {\bibfnamefont {W.}~\bibnamefont {Koshibae}}, \bibinfo {author}
  {\bibfnamefont {K.}~\bibnamefont {Ando}}, \bibinfo {author} {\bibfnamefont
  {S.}~\bibnamefont {Maekawa}},\ and\ \bibinfo {author} {\bibfnamefont
  {E.}~\bibnamefont {Saitoh}},\ }\href {https://doi.org/10.1038/nature07321}
  {\bibfield  {journal} {\bibinfo  {journal} {Nature (London)}\ }\textbf
  {\bibinfo {volume} {455}},\ \bibinfo {pages} {778} (\bibinfo {year}
  {2008})}\BibitemShut {NoStop}%
\bibitem [{\citenamefont {Jaworski}\ \emph {et~al.}(2010)\citenamefont
  {Jaworski}, \citenamefont {Yang}, \citenamefont {Mack}, \citenamefont
  {Awschalom}, \citenamefont {Heremans},\ and\ \citenamefont
  {Myers}}]{Jaworski2010}%
  \BibitemOpen
  \bibfield  {author} {\bibinfo {author} {\bibfnamefont {C.~M.}\ \bibnamefont
  {Jaworski}}, \bibinfo {author} {\bibfnamefont {J.}~\bibnamefont {Yang}},
  \bibinfo {author} {\bibfnamefont {S.}~\bibnamefont {Mack}}, \bibinfo {author}
  {\bibfnamefont {D.~D.}\ \bibnamefont {Awschalom}}, \bibinfo {author}
  {\bibfnamefont {J.~P.}\ \bibnamefont {Heremans}},\ and\ \bibinfo {author}
  {\bibfnamefont {R.~C.}\ \bibnamefont {Myers}},\ }\href
  {https://doi.org/10.1038/nmat2860} {\bibfield  {journal} {\bibinfo  {journal}
  {Nat. Mater.}\ }\textbf {\bibinfo {volume} {9}},\ \bibinfo {pages} {898}
  (\bibinfo {year} {2010})}\BibitemShut {NoStop}%
\bibitem [{\citenamefont {Bose}\ and\ \citenamefont
  {Tulapurkar}(2019)}]{BOSE2019165526}%
  \BibitemOpen
  \bibfield  {author} {\bibinfo {author} {\bibfnamefont {A.}~\bibnamefont
  {Bose}}\ and\ \bibinfo {author} {\bibfnamefont {A.~A.}\ \bibnamefont
  {Tulapurkar}},\ }\href {https://doi.org/10.1016/j.jmmm.2019.165526}
  {\bibfield  {journal} {\bibinfo  {journal} {J. Magn. Magn. Mater.}\ }\textbf
  {\bibinfo {volume} {491}},\ \bibinfo {pages} {165526} (\bibinfo {year}
  {2019})}\BibitemShut {NoStop}%
\bibitem [{\citenamefont {Sinova}\ \emph {et~al.}(2015)\citenamefont {Sinova},
  \citenamefont {Valenzuela}, \citenamefont {Wunderlich}, \citenamefont
  {Back},\ and\ \citenamefont {Jungwirth}}]{RevModPhys.87.1213}%
  \BibitemOpen
  \bibfield  {author} {\bibinfo {author} {\bibfnamefont {J.}~\bibnamefont
  {Sinova}}, \bibinfo {author} {\bibfnamefont {S.~O.}\ \bibnamefont
  {Valenzuela}}, \bibinfo {author} {\bibfnamefont {J.}~\bibnamefont
  {Wunderlich}}, \bibinfo {author} {\bibfnamefont {C.~H.}\ \bibnamefont
  {Back}},\ and\ \bibinfo {author} {\bibfnamefont {T.}~\bibnamefont
  {Jungwirth}},\ }\href {https://doi.org/10.1103/RevModPhys.87.1213} {\bibfield
   {journal} {\bibinfo  {journal} {Rev. Mod. Phys.}\ }\textbf {\bibinfo
  {volume} {87}},\ \bibinfo {pages} {1213} (\bibinfo {year}
  {2015})}\BibitemShut {NoStop}%
\bibitem [{\citenamefont {Cheng}\ \emph {et~al.}(2008)\citenamefont {Cheng},
  \citenamefont {Xing}, \citenamefont {Sun},\ and\ \citenamefont
  {Xie}}]{PhysRevB.78.045302}%
  \BibitemOpen
  \bibfield  {author} {\bibinfo {author} {\bibfnamefont {S.-g.}\ \bibnamefont
  {Cheng}}, \bibinfo {author} {\bibfnamefont {Y.}~\bibnamefont {Xing}},
  \bibinfo {author} {\bibfnamefont {Q.-f.}\ \bibnamefont {Sun}},\ and\ \bibinfo
  {author} {\bibfnamefont {X.~C.}\ \bibnamefont {Xie}},\ }\href
  {https://doi.org/10.1103/PhysRevB.78.045302} {\bibfield  {journal} {\bibinfo
  {journal} {Phys. Rev. B}\ }\textbf {\bibinfo {volume} {78}},\ \bibinfo
  {pages} {045302} (\bibinfo {year} {2008})}\BibitemShut {NoStop}%
\bibitem [{\citenamefont {Liu}\ and\ \citenamefont {Xie}(2010)}]{LIU2010471}%
  \BibitemOpen
  \bibfield  {author} {\bibinfo {author} {\bibfnamefont {X.}~\bibnamefont
  {Liu}}\ and\ \bibinfo {author} {\bibfnamefont {X.~C.}\ \bibnamefont {Xie}},\
  }\href {https://doi.org/10.1016/j.ssc.2009.12.017} {\bibfield  {journal}
  {\bibinfo  {journal} {Solid State Commun.}\ }\textbf {\bibinfo {volume}
  {150}},\ \bibinfo {pages} {471} (\bibinfo {year} {2010})}\BibitemShut
  {NoStop}%
\bibitem [{\citenamefont {Ma}(2010)}]{MA2010510}%
  \BibitemOpen
  \bibfield  {author} {\bibinfo {author} {\bibfnamefont {Z.}~\bibnamefont
  {Ma}},\ }\href {https://doi.org/10.1016/j.ssc.2009.11.004} {\bibfield
  {journal} {\bibinfo  {journal} {Solid State Commun.}\ }\textbf {\bibinfo
  {volume} {150}},\ \bibinfo {pages} {510} (\bibinfo {year}
  {2010})}\BibitemShut {NoStop}%
\bibitem [{\citenamefont {Meyer}\ \emph {et~al.}(2017)\citenamefont {Meyer},
  \citenamefont {Chen}, \citenamefont {Wimmer}, \citenamefont {Althammer},
  \citenamefont {Wimmer}, \citenamefont {Schlitz}, \citenamefont
  {S.~Gepr\"ags~and}, \citenamefont {K\"odderitzsch}, \citenamefont {Ebert},
  \citenamefont {Bauer}, \citenamefont {Gross},\ and\ \citenamefont
  {Goennenwein}}]{Meyer2017}%
  \BibitemOpen
  \bibfield  {author} {\bibinfo {author} {\bibfnamefont {S.}~\bibnamefont
  {Meyer}}, \bibinfo {author} {\bibfnamefont {Y.-T.}\ \bibnamefont {Chen}},
  \bibinfo {author} {\bibfnamefont {S.}~\bibnamefont {Wimmer}}, \bibinfo
  {author} {\bibfnamefont {M.}~\bibnamefont {Althammer}}, \bibinfo {author}
  {\bibfnamefont {T.}~\bibnamefont {Wimmer}}, \bibinfo {author} {\bibfnamefont
  {R.}~\bibnamefont {Schlitz}}, \bibinfo {author} {\bibfnamefont {H.~H.}\
  \bibnamefont {S.~Gepr\"ags~and}}, \bibinfo {author} {\bibfnamefont
  {D.}~\bibnamefont {K\"odderitzsch}}, \bibinfo {author} {\bibfnamefont
  {H.}~\bibnamefont {Ebert}}, \bibinfo {author} {\bibfnamefont {G.~E.~W.}\
  \bibnamefont {Bauer}}, \bibinfo {author} {\bibfnamefont {R.}~\bibnamefont
  {Gross}},\ and\ \bibinfo {author} {\bibfnamefont {S.~T.~B.}\ \bibnamefont
  {Goennenwein}},\ }\href {https://doi.org/10.1038/nmat4964} {\bibfield
  {journal} {\bibinfo  {journal} {Nat. Mater.}\ }\textbf {\bibinfo {volume}
  {16}},\ \bibinfo {pages} {977} (\bibinfo {year} {2017})}\BibitemShut
  {NoStop}%
\bibitem [{\citenamefont {Sheng}\ \emph {et~al.}(2017)\citenamefont {Sheng},
  \citenamefont {Sakuraba}, \citenamefont {Lau}, \citenamefont {Takahashi},
  \citenamefont {Mitani},\ and\ \citenamefont {Hayashi}}]{Shenge1701503}%
  \BibitemOpen
  \bibfield  {author} {\bibinfo {author} {\bibfnamefont {P.}~\bibnamefont
  {Sheng}}, \bibinfo {author} {\bibfnamefont {Y.}~\bibnamefont {Sakuraba}},
  \bibinfo {author} {\bibfnamefont {Y.-C.}\ \bibnamefont {Lau}}, \bibinfo
  {author} {\bibfnamefont {S.}~\bibnamefont {Takahashi}}, \bibinfo {author}
  {\bibfnamefont {S.}~\bibnamefont {Mitani}},\ and\ \bibinfo {author}
  {\bibfnamefont {M.}~\bibnamefont {Hayashi}},\ }\href
  {https://doi.org/10.1126/sciadv.1701503} {\bibfield  {journal} {\bibinfo
  {journal} {Sci. Adv.}\ }\textbf {\bibinfo {volume} {3}},\ \bibinfo {pages}
  {e1701503} (\bibinfo {year} {2017})}\BibitemShut {NoStop}%
\bibitem [{\citenamefont {Kim}\ \emph {et~al.}(2017)\citenamefont {Kim},
  \citenamefont {Jeon}, \citenamefont {Choi}, \citenamefont {Lee},
  \citenamefont {Surabhi}, \citenamefont {Jeong}, \citenamefont {Lee},\ and\
  \citenamefont {Park}}]{ncomms1400}%
  \BibitemOpen
  \bibfield  {author} {\bibinfo {author} {\bibfnamefont {D.-J.}\ \bibnamefont
  {Kim}}, \bibinfo {author} {\bibfnamefont {C.-Y.}\ \bibnamefont {Jeon}},
  \bibinfo {author} {\bibfnamefont {J.-G.}\ \bibnamefont {Choi}}, \bibinfo
  {author} {\bibfnamefont {J.~W.}\ \bibnamefont {Lee}}, \bibinfo {author}
  {\bibfnamefont {S.}~\bibnamefont {Surabhi}}, \bibinfo {author} {\bibfnamefont
  {J.-R.}\ \bibnamefont {Jeong}}, \bibinfo {author} {\bibfnamefont {K.-J.}\
  \bibnamefont {Lee}},\ and\ \bibinfo {author} {\bibfnamefont {B.-G.}\
  \bibnamefont {Park}},\ }\href {https://doi.org/10.1038/s41467-017-01493-5}
  {\bibfield  {journal} {\bibinfo  {journal} {Nat. Commun.}\ }\textbf {\bibinfo
  {volume} {8}},\ \bibinfo {pages} {1400} (\bibinfo {year} {2017})}\BibitemShut
  {NoStop}%
\bibitem [{\citenamefont {Sheng}\ \emph {et~al.}(2005)\citenamefont {Sheng},
  \citenamefont {Sheng},\ and\ \citenamefont {Ting}}]{PhysRevLett.94.016602}%
  \BibitemOpen
  \bibfield  {author} {\bibinfo {author} {\bibfnamefont {L.}~\bibnamefont
  {Sheng}}, \bibinfo {author} {\bibfnamefont {D.~N.}\ \bibnamefont {Sheng}},\
  and\ \bibinfo {author} {\bibfnamefont {C.~S.}\ \bibnamefont {Ting}},\ }\href
  {https://doi.org/10.1103/PhysRevLett.94.016602} {\bibfield  {journal}
  {\bibinfo  {journal} {Phys. Rev. Lett.}\ }\textbf {\bibinfo {volume} {94}},\
  \bibinfo {pages} {016602} (\bibinfo {year} {2005})}\BibitemShut {NoStop}%
\bibitem [{\citenamefont {Nikoli\'{c}}\ \emph {et~al.}(2005)\citenamefont
  {Nikoli\'{c}}, \citenamefont {Z\^arbo},\ and\ \citenamefont
  {Souma}}]{PhysRevB.72.075361}%
  \BibitemOpen
  \bibfield  {author} {\bibinfo {author} {\bibfnamefont {B.~K.}\ \bibnamefont
  {Nikoli\'{c}}}, \bibinfo {author} {\bibfnamefont {L.~P.}\ \bibnamefont
  {Z\^arbo}},\ and\ \bibinfo {author} {\bibfnamefont {S.}~\bibnamefont
  {Souma}},\ }\href {https://doi.org/10.1103/PhysRevB.72.075361} {\bibfield
  {journal} {\bibinfo  {journal} {Phys. Rev. B}\ }\textbf {\bibinfo {volume}
  {72}},\ \bibinfo {pages} {075361} (\bibinfo {year} {2005})}\BibitemShut
  {NoStop}%
\bibitem [{\citenamefont {Rashba}(2003)}]{PhysRevB.68.241315}%
  \BibitemOpen
  \bibfield  {author} {\bibinfo {author} {\bibfnamefont {E.~I.}\ \bibnamefont
  {Rashba}},\ }\href {https://doi.org/10.1103/PhysRevB.68.241315} {\bibfield
  {journal} {\bibinfo  {journal} {Phys. Rev. B}\ }\textbf {\bibinfo {volume}
  {68}},\ \bibinfo {pages} {241315(R)} (\bibinfo {year} {2003})}\BibitemShut
  {NoStop}%
\bibitem [{\citenamefont {Shi}\ \emph {et~al.}(2006)\citenamefont {Shi},
  \citenamefont {Zhang}, \citenamefont {Xiao},\ and\ \citenamefont
  {Niu}}]{PhysRevLett.96.076604}%
  \BibitemOpen
  \bibfield  {author} {\bibinfo {author} {\bibfnamefont {J.}~\bibnamefont
  {Shi}}, \bibinfo {author} {\bibfnamefont {P.}~\bibnamefont {Zhang}}, \bibinfo
  {author} {\bibfnamefont {D.}~\bibnamefont {Xiao}},\ and\ \bibinfo {author}
  {\bibfnamefont {Q.}~\bibnamefont {Niu}},\ }\href
  {https://doi.org/10.1103/PhysRevLett.96.076604} {\bibfield  {journal}
  {\bibinfo  {journal} {Phys. Rev. Lett.}\ }\textbf {\bibinfo {volume} {96}},\
  \bibinfo {pages} {076604} (\bibinfo {year} {2006})}\BibitemShut {NoStop}%
\bibitem [{\citenamefont {Zhang}\ \emph {et~al.}(2008)\citenamefont {Zhang},
  \citenamefont {Wang}, \citenamefont {Shi}, \citenamefont {Xiao},\ and\
  \citenamefont {Niu}}]{PhysRevB.77.075304}%
  \BibitemOpen
  \bibfield  {author} {\bibinfo {author} {\bibfnamefont {P.}~\bibnamefont
  {Zhang}}, \bibinfo {author} {\bibfnamefont {Z.}~\bibnamefont {Wang}},
  \bibinfo {author} {\bibfnamefont {J.}~\bibnamefont {Shi}}, \bibinfo {author}
  {\bibfnamefont {D.}~\bibnamefont {Xiao}},\ and\ \bibinfo {author}
  {\bibfnamefont {Q.}~\bibnamefont {Niu}},\ }\href
  {https://doi.org/10.1103/PhysRevB.77.075304} {\bibfield  {journal} {\bibinfo
  {journal} {Phys. Rev. B}\ }\textbf {\bibinfo {volume} {77}},\ \bibinfo
  {pages} {075304} (\bibinfo {year} {2008})}\BibitemShut {NoStop}%
\bibitem [{\citenamefont {Smrcka}\ and\ \citenamefont
  {Streda}(1977)}]{0022-3719-10-12-021}%
  \BibitemOpen
  \bibfield  {author} {\bibinfo {author} {\bibfnamefont {L.}~\bibnamefont
  {Smrcka}}\ and\ \bibinfo {author} {\bibfnamefont {P.}~\bibnamefont
  {Streda}},\ }\href {https://doi.org/10.1088/0022-3719/10/12/021}
  {\bibfield  {journal} {\bibinfo  {journal} {J. Phys. C: Solid State Phys.}\
  }\textbf {\bibinfo {volume} {10}},\ \bibinfo {pages} {2153} (\bibinfo {year}
  {1977})}\BibitemShut {NoStop}%
\bibitem [{\citenamefont {Cooper}\ \emph {et~al.}(1997)\citenamefont {Cooper},
  \citenamefont {Halperin},\ and\ \citenamefont {Ruzin}}]{PhysRevB.55.2344}%
  \BibitemOpen
  \bibfield  {author} {\bibinfo {author} {\bibfnamefont {N.~R.}\ \bibnamefont
  {Cooper}}, \bibinfo {author} {\bibfnamefont {B.~I.}\ \bibnamefont
  {Halperin}},\ and\ \bibinfo {author} {\bibfnamefont {I.~M.}\ \bibnamefont
  {Ruzin}},\ }\href {https://doi.org/10.1103/PhysRevB.55.2344} {\bibfield
  {journal} {\bibinfo  {journal} {Phys. Rev. B}\ }\textbf {\bibinfo {volume}
  {55}},\ \bibinfo {pages} {2344} (\bibinfo {year} {1997})}\BibitemShut
  {NoStop}%
\bibitem [{\citenamefont {Matsumoto}\ and\ \citenamefont
  {Murakami}(2011{\natexlab{a}})}]{PhysRevLett.106.197202}%
  \BibitemOpen
  \bibfield  {author} {\bibinfo {author} {\bibfnamefont {R.}~\bibnamefont
  {Matsumoto}}\ and\ \bibinfo {author} {\bibfnamefont {S.}~\bibnamefont
  {Murakami}},\ }\href {https://doi.org/10.1103/PhysRevLett.106.197202}
  {\bibfield  {journal} {\bibinfo  {journal} {Phys. Rev. Lett.}\ }\textbf
  {\bibinfo {volume} {106}},\ \bibinfo {pages} {197202} (\bibinfo {year}
  {2011}{\natexlab{a}})}\BibitemShut {NoStop}%
\bibitem [{\citenamefont {Matsumoto}\ and\ \citenamefont
  {Murakami}(2011{\natexlab{b}})}]{PhysRevB.84.184406}%
  \BibitemOpen
  \bibfield  {author} {\bibinfo {author} {\bibfnamefont {R.}~\bibnamefont
  {Matsumoto}}\ and\ \bibinfo {author} {\bibfnamefont {S.}~\bibnamefont
  {Murakami}},\ }\href {https://doi.org/10.1103/PhysRevB.84.184406} {\bibfield
  {journal} {\bibinfo  {journal} {Phys. Rev. B}\ }\textbf {\bibinfo {volume}
  {84}},\ \bibinfo {pages} {184406} (\bibinfo {year}
  {2011}{\natexlab{b}})}\BibitemShut {NoStop}%
\bibitem [{\citenamefont {Qin}\ \emph {et~al.}(2011)\citenamefont {Qin},
  \citenamefont {Niu},\ and\ \citenamefont {Shi}}]{PhysRevLett.107.236601}%
  \BibitemOpen
  \bibfield  {author} {\bibinfo {author} {\bibfnamefont {T.}~\bibnamefont
  {Qin}}, \bibinfo {author} {\bibfnamefont {Q.}~\bibnamefont {Niu}},\ and\
  \bibinfo {author} {\bibfnamefont {J.}~\bibnamefont {Shi}},\ }\href
  {https://doi.org/10.1103/PhysRevLett.107.236601} {\bibfield  {journal}
  {\bibinfo  {journal} {Phys. Rev. Lett.}\ }\textbf {\bibinfo {volume} {107}},\
  \bibinfo {pages} {236601} (\bibinfo {year} {2011})}\BibitemShut {NoStop}%
\bibitem [{\citenamefont {Luttinger}(1964)}]{PhysRev.135.A1505}%
  \BibitemOpen
  \bibfield  {author} {\bibinfo {author} {\bibfnamefont {J.~M.}\ \bibnamefont
  {Luttinger}},\ }\href {https://doi.org/10.1103/PhysRev.135.A1505} {\bibfield
  {journal} {\bibinfo  {journal} {Phys. Rev.}\ }\textbf {\bibinfo {volume}
  {135}},\ \bibinfo {pages} {A1505} (\bibinfo {year} {1964})}\BibitemShut
  {NoStop}%
\bibitem [{\citenamefont {Shitade}\ \emph {et~al.}(2019)\citenamefont
  {Shitade}, \citenamefont {Daido},\ and\ \citenamefont
  {Yanase}}]{PhysRevB.99.024404}%
  \BibitemOpen
  \bibfield  {author} {\bibinfo {author} {\bibfnamefont {A.}~\bibnamefont
  {Shitade}}, \bibinfo {author} {\bibfnamefont {A.}~\bibnamefont {Daido}},\
  and\ \bibinfo {author} {\bibfnamefont {Y.}~\bibnamefont {Yanase}},\ }\href
  {https://doi.org/10.1103/PhysRevB.99.024404} {\bibfield  {journal} {\bibinfo
  {journal} {Phys. Rev. B}\ }\textbf {\bibinfo {volume} {99}},\ \bibinfo
  {pages} {024404} (\bibinfo {year} {2019})}\BibitemShut {NoStop}%
\bibitem [{\citenamefont {Dong}\ \emph {et~al.}(2020)\citenamefont {Dong},
  \citenamefont {Xiao}, \citenamefont {Xiong},\ and\ \citenamefont
  {Niu}}]{PhysRevLett.124.066601}%
  \BibitemOpen
  \bibfield  {author} {\bibinfo {author} {\bibfnamefont {L.}~\bibnamefont
  {Dong}}, \bibinfo {author} {\bibfnamefont {C.}~\bibnamefont {Xiao}}, \bibinfo
  {author} {\bibfnamefont {B.}~\bibnamefont {Xiong}},\ and\ \bibinfo {author}
  {\bibfnamefont {Q.}~\bibnamefont {Niu}},\ }\href
  {https://doi.org/10.1103/PhysRevLett.124.066601} {\bibfield  {journal}
  {\bibinfo  {journal} {Phys. Rev. Lett.}\ }\textbf {\bibinfo {volume} {124}},\
  \bibinfo {pages} {066601} (\bibinfo {year} {2020})}\BibitemShut {NoStop}%
\bibitem [{\citenamefont {Wimmer}\ \emph {et~al.}(2013)\citenamefont {Wimmer},
  \citenamefont {K\"odderitzsch}, \citenamefont {Chadova},\ and\ \citenamefont
  {Ebert}}]{PhysRevB.88.201108}%
  \BibitemOpen
  \bibfield  {author} {\bibinfo {author} {\bibfnamefont {S.}~\bibnamefont
  {Wimmer}}, \bibinfo {author} {\bibfnamefont {D.}~\bibnamefont
  {K\"odderitzsch}}, \bibinfo {author} {\bibfnamefont {K.}~\bibnamefont
  {Chadova}},\ and\ \bibinfo {author} {\bibfnamefont {H.}~\bibnamefont
  {Ebert}},\ }\href {https://doi.org/10.1103/PhysRevB.88.201108} {\bibfield
  {journal} {\bibinfo  {journal} {Phys. Rev. B}\ }\textbf {\bibinfo {volume}
  {88}},\ \bibinfo {pages} {201108(R)} (\bibinfo {year} {2013})}\BibitemShut
  {NoStop}%
\bibitem [{\citenamefont {Guo}\ and\ \citenamefont
  {Wang}(2017)}]{PhysRevB.96.224415}%
  \BibitemOpen
  \bibfield  {author} {\bibinfo {author} {\bibfnamefont {G.-Y.}\ \bibnamefont
  {Guo}}\ and\ \bibinfo {author} {\bibfnamefont {T.-C.}\ \bibnamefont {Wang}},\
  }\href {https://doi.org/10.1103/PhysRevB.96.224415} {\bibfield  {journal}
  {\bibinfo  {journal} {Phys. Rev. B}\ }\textbf {\bibinfo {volume} {96}},\
  \bibinfo {pages} {224415} (\bibinfo {year} {2017})}\BibitemShut {NoStop}%
\bibitem [{\citenamefont {Guo}\ and\ \citenamefont
  {Wang}(2019)}]{PhysRevB.100.169907}%
  \BibitemOpen
  \bibfield  {author} {\bibinfo {author} {\bibfnamefont {G.-Y.}\ \bibnamefont
  {Guo}}\ and\ \bibinfo {author} {\bibfnamefont {T.-C.}\ \bibnamefont {Wang}},\
  }\href {https://doi.org/10.1103/PhysRevB.100.169907} {\bibfield  {journal}
  {\bibinfo  {journal} {Phys. Rev. B}\ }\textbf {\bibinfo {volume} {100}},\
  \bibinfo {pages} {169907(E)} (\bibinfo {year} {2019})}\BibitemShut {NoStop}%
\bibitem [{\citenamefont {Yen}\ and\ \citenamefont
  {Guo}(2020)}]{PhysRevB.101.064430}%
  \BibitemOpen
  \bibfield  {author} {\bibinfo {author} {\bibfnamefont {Y.}~\bibnamefont
  {Yen}}\ and\ \bibinfo {author} {\bibfnamefont {G.-Y.}\ \bibnamefont {Guo}},\
  }\href {https://doi.org/10.1103/PhysRevB.101.064430} {\bibfield  {journal}
  {\bibinfo  {journal} {Phys. Rev. B}\ }\textbf {\bibinfo {volume} {101}},\
  \bibinfo {pages} {064430} (\bibinfo {year} {2020})}\BibitemShut {NoStop}%
\bibitem [{\citenamefont {Zhang}\ \emph {et~al.}(2020)\citenamefont {Zhang},
  \citenamefont {Xu}, \citenamefont {Koepernik}, \citenamefont {Fu},
  \citenamefont {Gooth}, \citenamefont {van~den Brink}, \citenamefont
  {Felser},\ and\ \citenamefont {Sun}}]{Zhang_2020}%
  \BibitemOpen
  \bibfield  {author} {\bibinfo {author} {\bibfnamefont {Y.}~\bibnamefont
  {Zhang}}, \bibinfo {author} {\bibfnamefont {Q.}~\bibnamefont {Xu}}, \bibinfo
  {author} {\bibfnamefont {K.}~\bibnamefont {Koepernik}}, \bibinfo {author}
  {\bibfnamefont {C.}~\bibnamefont {Fu}}, \bibinfo {author} {\bibfnamefont
  {J.}~\bibnamefont {Gooth}}, \bibinfo {author} {\bibfnamefont
  {J.}~\bibnamefont {van~den Brink}}, \bibinfo {author} {\bibfnamefont
  {C.}~\bibnamefont {Felser}},\ and\ \bibinfo {author} {\bibfnamefont
  {Y.}~\bibnamefont {Sun}},\ }\href {https://doi.org/10.1088/1367-2630/abaa87}
  {\bibfield  {journal} {\bibinfo  {journal} {New J. Phys.}\ }\textbf {\bibinfo
  {volume} {22}},\ \bibinfo {pages} {093003} (\bibinfo {year}
  {2020})}\BibitemShut {NoStop}%
\bibitem [{\citenamefont {Prasad}\ and\ \citenamefont
  {Guo}(2020)}]{PhysRevMaterials.4.124205}%
  \BibitemOpen
  \bibfield  {author} {\bibinfo {author} {\bibfnamefont {B.~B.}\ \bibnamefont
  {Prasad}}\ and\ \bibinfo {author} {\bibfnamefont {G.-Y.}\ \bibnamefont
  {Guo}},\ }\href {https://doi.org/10.1103/PhysRevMaterials.4.124205}
  {\bibfield  {journal} {\bibinfo  {journal} {Phys. Rev. Mater.}\ }\textbf
  {\bibinfo {volume} {4}},\ \bibinfo {pages} {124205} (\bibinfo {year}
  {2020})}\BibitemShut {NoStop}%
\bibitem [{\citenamefont {Dyrda\l}\ \emph
  {et~al.}(2016{\natexlab{a}})\citenamefont {Dyrda\l}, \citenamefont
  {Barna\'{s}},\ and\ \citenamefont {Dugaev}}]{PhysRevB.94.035306}%
  \BibitemOpen
  \bibfield  {author} {\bibinfo {author} {\bibfnamefont {A.}~\bibnamefont
  {Dyrda\l}}, \bibinfo {author} {\bibfnamefont {J.}~\bibnamefont
  {Barna\'{s}}},\ and\ \bibinfo {author} {\bibfnamefont {V.~K.}\ \bibnamefont
  {Dugaev}},\ }\href {https://doi.org/10.1103/PhysRevB.94.035306} {\bibfield
  {journal} {\bibinfo  {journal} {Phys. Rev. B}\ }\textbf {\bibinfo {volume}
  {94}},\ \bibinfo {pages} {035306} (\bibinfo {year}
  {2016}{\natexlab{a}})}\BibitemShut {NoStop}%
\bibitem [{\citenamefont {Dyrda\l}\ \emph
  {et~al.}(2016{\natexlab{b}})\citenamefont {Dyrda\l}, \citenamefont {Dugaev},\
  and\ \citenamefont {Barna\'{s}}}]{PhysRevB.94.205302}%
  \BibitemOpen
  \bibfield  {author} {\bibinfo {author} {\bibfnamefont {A.}~\bibnamefont
  {Dyrda\l}}, \bibinfo {author} {\bibfnamefont {V.~K.}\ \bibnamefont
  {Dugaev}},\ and\ \bibinfo {author} {\bibfnamefont {J.}~\bibnamefont
  {Barna\'{s}}},\ }\href {https://doi.org/10.1103/PhysRevB.94.205302}
  {\bibfield  {journal} {\bibinfo  {journal} {Phys. Rev. B}\ }\textbf {\bibinfo
  {volume} {94}},\ \bibinfo {pages} {205302} (\bibinfo {year}
  {2016}{\natexlab{b}})}\BibitemShut {NoStop}%
\bibitem [{\citenamefont {Jiang}\ and\ \citenamefont
  {Ma}(2022)}]{PhysRevB.105.035302}%
  \BibitemOpen
  \bibfield  {author} {\bibinfo {author} {\bibfnamefont {P.}~\bibnamefont
  {Jiang}}\ and\ \bibinfo {author} {\bibfnamefont {Z.}~\bibnamefont {Ma}},\
  }\href {https://doi.org/10.1103/PhysRevB.105.035302} {\bibfield  {journal}
  {\bibinfo  {journal} {Phys. Rev. B}\ }\textbf {\bibinfo {volume} {105}},\
  \bibinfo {pages} {035302} (\bibinfo {year} {2022})}\BibitemShut {NoStop}%
\bibitem [{\citenamefont {Xiao}\ \emph {et~al.}(2018)\citenamefont {Xiao},
  \citenamefont {Zhu}, \citenamefont {Xiong},\ and\ \citenamefont
  {Niu}}]{PhysRevB.98.081401}%
  \BibitemOpen
  \bibfield  {author} {\bibinfo {author} {\bibfnamefont {C.}~\bibnamefont
  {Xiao}}, \bibinfo {author} {\bibfnamefont {J.}~\bibnamefont {Zhu}}, \bibinfo
  {author} {\bibfnamefont {B.}~\bibnamefont {Xiong}},\ and\ \bibinfo {author}
  {\bibfnamefont {Q.}~\bibnamefont {Niu}},\ }\href
  {https://doi.org/10.1103/PhysRevB.98.081401} {\bibfield  {journal} {\bibinfo
  {journal} {Phys. Rev. B}\ }\textbf {\bibinfo {volume} {98}},\ \bibinfo
  {pages} {081401(R)} (\bibinfo {year} {2018})}\BibitemShut {NoStop}%
\bibitem [{\citenamefont {Xiao}\ and\ \citenamefont
  {Niu}(2021)}]{PhysRevB.104.L241411}%
  \BibitemOpen
  \bibfield  {author} {\bibinfo {author} {\bibfnamefont {C.}~\bibnamefont
  {Xiao}}\ and\ \bibinfo {author} {\bibfnamefont {Q.}~\bibnamefont {Niu}},\
  }\href {https://doi.org/10.1103/PhysRevB.104.L241411} {\bibfield  {journal}
  {\bibinfo  {journal} {Phys. Rev. B}\ }\textbf {\bibinfo {volume} {104}},\
  \bibinfo {pages} {L241411} (\bibinfo {year} {2021})}\BibitemShut {NoStop}%
\bibitem [{\citenamefont {D'yakonov}\ and\ \citenamefont
  {Perel'}(1971)}]{Dyakonov1971}%
  \BibitemOpen
  \bibfield  {author} {\bibinfo {author} {\bibfnamefont {M.~I.}\ \bibnamefont
  {D'yakonov}}\ and\ \bibinfo {author} {\bibfnamefont {V.~I.}\ \bibnamefont
  {Perel'}},\ }\href@noop {} {\bibfield  {journal} {\bibinfo  {journal} {ZhETF
  Pis'ma Red.}\ }\textbf {\bibinfo {volume} {13}},\ \bibinfo {pages} {657}
  (\bibinfo {year} {1971})},\
  \translation{\href{http://jetpletters.ru/ps/1587/article_24366.shtml}{JETP
  Lett. \textbf{13}, 467 (1971)}}\BibitemShut {NoStop}%
\bibitem [{\citenamefont {Ma}\ \emph {et~al.}(2004)\citenamefont {Ma},
  \citenamefont {Hu}, \citenamefont {Tao},\ and\ \citenamefont
  {Shen}}]{PhysRevB.70.195343}%
  \BibitemOpen
  \bibfield  {author} {\bibinfo {author} {\bibfnamefont {X.}~\bibnamefont
  {Ma}}, \bibinfo {author} {\bibfnamefont {L.}~\bibnamefont {Hu}}, \bibinfo
  {author} {\bibfnamefont {R.}~\bibnamefont {Tao}},\ and\ \bibinfo {author}
  {\bibfnamefont {S.-Q.}\ \bibnamefont {Shen}},\ }\href
  {https://doi.org/10.1103/PhysRevB.70.195343} {\bibfield  {journal} {\bibinfo
  {journal} {Phys. Rev. B}\ }\textbf {\bibinfo {volume} {70}},\ \bibinfo
  {pages} {195343} (\bibinfo {year} {2004})}\BibitemShut {NoStop}%
\bibitem [{\citenamefont {Nikoli\'c}\ \emph {et~al.}(2005)\citenamefont
  {Nikoli\'c}, \citenamefont {Souma}, \citenamefont {Z\^arbo},\ and\
  \citenamefont {Sinova}}]{PhysRevLett.95.046601}%
  \BibitemOpen
  \bibfield  {author} {\bibinfo {author} {\bibfnamefont {B.~K.}\ \bibnamefont
  {Nikoli\'c}}, \bibinfo {author} {\bibfnamefont {S.}~\bibnamefont {Souma}},
  \bibinfo {author} {\bibfnamefont {L.~P.}\ \bibnamefont {Z\^arbo}},\ and\
  \bibinfo {author} {\bibfnamefont {J.}~\bibnamefont {Sinova}},\ }\href
  {https://doi.org/10.1103/PhysRevLett.95.046601} {\bibfield  {journal}
  {\bibinfo  {journal} {Phys. Rev. Lett.}\ }\textbf {\bibinfo {volume} {95}},\
  \bibinfo {pages} {046601} (\bibinfo {year} {2005})}\BibitemShut {NoStop}%
\bibitem [{\citenamefont {Onoda}\ and\ \citenamefont
  {Nagaosa}(2005)}]{PhysRevB.72.081301}%
  \BibitemOpen
  \bibfield  {author} {\bibinfo {author} {\bibfnamefont {M.}~\bibnamefont
  {Onoda}}\ and\ \bibinfo {author} {\bibfnamefont {N.}~\bibnamefont
  {Nagaosa}},\ }\href {https://doi.org/10.1103/PhysRevB.72.081301} {\bibfield
  {journal} {\bibinfo  {journal} {Phys. Rev. B}\ }\textbf {\bibinfo {volume}
  {72}},\ \bibinfo {pages} {081301(R)} (\bibinfo {year} {2005})}\BibitemShut
  {NoStop}%
\bibitem [{\citenamefont {Nikoli\'c}\ \emph {et~al.}(2006)\citenamefont
  {Nikoli\'c}, \citenamefont {Z\^arbo},\ and\ \citenamefont
  {Souma}}]{PhysRevB.73.075303}%
  \BibitemOpen
  \bibfield  {author} {\bibinfo {author} {\bibfnamefont {B.~K.}\ \bibnamefont
  {Nikoli\'c}}, \bibinfo {author} {\bibfnamefont {L.~P.}\ \bibnamefont
  {Z\^arbo}},\ and\ \bibinfo {author} {\bibfnamefont {S.}~\bibnamefont
  {Souma}},\ }\href {https://doi.org/10.1103/PhysRevB.73.075303} {\bibfield
  {journal} {\bibinfo  {journal} {Phys. Rev. B}\ }\textbf {\bibinfo {volume}
  {73}},\ \bibinfo {pages} {075303} (\bibinfo {year} {2006})}\BibitemShut
  {NoStop}%
\bibitem [{\citenamefont {Rashba}(2006)}]{RASHBA200631}%
  \BibitemOpen
  \bibfield  {author} {\bibinfo {author} {\bibfnamefont {E.~I.}\ \bibnamefont
  {Rashba}},\ }\href {https://doi.org/10.1016/j.physe.2006.02.014} {\bibfield
  {journal} {\bibinfo  {journal} {Physica E}\ }\textbf {\bibinfo {volume}
  {34}},\ \bibinfo {pages} {31} (\bibinfo {year} {2006})}\BibitemShut {NoStop}%
\bibitem [{\citenamefont {Raimondi}\ \emph {et~al.}(2006)\citenamefont
  {Raimondi}, \citenamefont {Gorini}, \citenamefont {Schwab},\ and\
  \citenamefont {Dzierzawa}}]{PhysRevB.74.035340}%
  \BibitemOpen
  \bibfield  {author} {\bibinfo {author} {\bibfnamefont {R.}~\bibnamefont
  {Raimondi}}, \bibinfo {author} {\bibfnamefont {C.}~\bibnamefont {Gorini}},
  \bibinfo {author} {\bibfnamefont {P.}~\bibnamefont {Schwab}},\ and\ \bibinfo
  {author} {\bibfnamefont {M.}~\bibnamefont {Dzierzawa}},\ }\href
  {https://doi.org/10.1103/PhysRevB.74.035340} {\bibfield  {journal} {\bibinfo
  {journal} {Phys. Rev. B}\ }\textbf {\bibinfo {volume} {74}},\ \bibinfo
  {pages} {035340} (\bibinfo {year} {2006})}\BibitemShut {NoStop}%
\bibitem [{\citenamefont {Zyuzin}\ \emph {et~al.}(2007)\citenamefont {Zyuzin},
  \citenamefont {Silvestrov},\ and\ \citenamefont
  {Mishchenko}}]{PhysRevLett.99.106601}%
  \BibitemOpen
  \bibfield  {author} {\bibinfo {author} {\bibfnamefont {V.~A.}\ \bibnamefont
  {Zyuzin}}, \bibinfo {author} {\bibfnamefont {P.~G.}\ \bibnamefont
  {Silvestrov}},\ and\ \bibinfo {author} {\bibfnamefont {E.~G.}\ \bibnamefont
  {Mishchenko}},\ }\href {https://doi.org/10.1103/PhysRevLett.99.106601}
  {\bibfield  {journal} {\bibinfo  {journal} {Phys. Rev. Lett.}\ }\textbf
  {\bibinfo {volume} {99}},\ \bibinfo {pages} {106601} (\bibinfo {year}
  {2007})}\BibitemShut {NoStop}%
\bibitem [{\citenamefont {Silvestrov}\ \emph {et~al.}(2009)\citenamefont
  {Silvestrov}, \citenamefont {Zyuzin},\ and\ \citenamefont
  {Mishchenko}}]{PhysRevLett.102.196802}%
  \BibitemOpen
  \bibfield  {author} {\bibinfo {author} {\bibfnamefont {P.~G.}\ \bibnamefont
  {Silvestrov}}, \bibinfo {author} {\bibfnamefont {V.~A.}\ \bibnamefont
  {Zyuzin}},\ and\ \bibinfo {author} {\bibfnamefont {E.~G.}\ \bibnamefont
  {Mishchenko}},\ }\href {https://doi.org/10.1103/PhysRevLett.102.196802}
  {\bibfield  {journal} {\bibinfo  {journal} {Phys. Rev. Lett.}\ }\textbf
  {\bibinfo {volume} {102}},\ \bibinfo {pages} {196802} (\bibinfo {year}
  {2009})}\BibitemShut {NoStop}%
\bibitem [{\citenamefont {Sonin}(2010)}]{PhysRevB.81.113304}%
  \BibitemOpen
  \bibfield  {author} {\bibinfo {author} {\bibfnamefont {E.~B.}\ \bibnamefont
  {Sonin}},\ }\href {https://doi.org/10.1103/PhysRevB.81.113304} {\bibfield
  {journal} {\bibinfo  {journal} {Phys. Rev. B}\ }\textbf {\bibinfo {volume}
  {81}},\ \bibinfo {pages} {113304} (\bibinfo {year} {2010})}\BibitemShut
  {NoStop}%
\bibitem [{\citenamefont {Hosono}\ \emph {et~al.}(2011)\citenamefont {Hosono},
  \citenamefont {Yamaguchi}, \citenamefont {Nozaki},\ and\ \citenamefont
  {Tatara}}]{PhysRevB.83.144428}%
  \BibitemOpen
  \bibfield  {author} {\bibinfo {author} {\bibfnamefont {K.}~\bibnamefont
  {Hosono}}, \bibinfo {author} {\bibfnamefont {A.}~\bibnamefont {Yamaguchi}},
  \bibinfo {author} {\bibfnamefont {Y.}~\bibnamefont {Nozaki}},\ and\ \bibinfo
  {author} {\bibfnamefont {G.}~\bibnamefont {Tatara}},\ }\href
  {https://doi.org/10.1103/PhysRevB.83.144428} {\bibfield  {journal} {\bibinfo
  {journal} {Phys. Rev. B}\ }\textbf {\bibinfo {volume} {83}},\ \bibinfo
  {pages} {144428} (\bibinfo {year} {2011})}\BibitemShut {NoStop}%
\bibitem [{\citenamefont {Tatara}(2018)}]{PhysRevB.98.174422}%
  \BibitemOpen
  \bibfield  {author} {\bibinfo {author} {\bibfnamefont {G.}~\bibnamefont
  {Tatara}},\ }\href {https://doi.org/10.1103/PhysRevB.98.174422} {\bibfield
  {journal} {\bibinfo  {journal} {Phys. Rev. B}\ }\textbf {\bibinfo {volume}
  {98}},\ \bibinfo {pages} {174422} (\bibinfo {year} {2018})}\BibitemShut
  {NoStop}%
\bibitem [{\citenamefont {Shitade}\ and\ \citenamefont
  {Tatara}(2022)}]{PhysRevB.105.L201202}%
  \BibitemOpen
  \bibfield  {author} {\bibinfo {author} {\bibfnamefont {A.}~\bibnamefont
  {Shitade}}\ and\ \bibinfo {author} {\bibfnamefont {G.}~\bibnamefont
  {Tatara}},\ }\href {https://doi.org/10.1103/PhysRevB.105.L201202} {\bibfield
  {journal} {\bibinfo  {journal} {Phys. Rev. B}\ }\textbf {\bibinfo {volume}
  {105}},\ \bibinfo {pages} {L201202} (\bibinfo {year} {2022})}\BibitemShut
  {NoStop}%
\bibitem [{\citenamefont {Luttinger}(1956)}]{PhysRev.102.1030}%
  \BibitemOpen
  \bibfield  {author} {\bibinfo {author} {\bibfnamefont {J.~M.}\ \bibnamefont
  {Luttinger}},\ }\href {https://doi.org/10.1103/PhysRev.102.1030} {\bibfield
  {journal} {\bibinfo  {journal} {Phys. Rev.}\ }\textbf {\bibinfo {volume}
  {102}},\ \bibinfo {pages} {1030} (\bibinfo {year} {1956})}\BibitemShut
  {NoStop}%
\bibitem [{\citenamefont {Shitade}(2014)}]{Shitade01122014}%
  \BibitemOpen
  \bibfield  {author} {\bibinfo {author} {\bibfnamefont {A.}~\bibnamefont
  {Shitade}},\ }\href {https://doi.org/10.1093/ptep/ptu162} {\bibfield
  {journal} {\bibinfo  {journal} {Prog. Theor. Exp. Phys.}\ }\textbf {\bibinfo
  {volume} {2014}},\ \bibinfo {pages} {123I01} (\bibinfo {year}
  {2014})}\BibitemShut {NoStop}%
\bibitem [{\citenamefont {Shitade}(2017)}]{JPSJ.86.054601}%
  \BibitemOpen
  \bibfield  {author} {\bibinfo {author} {\bibfnamefont {A.}~\bibnamefont
  {Shitade}},\ }\href {https://doi.org/10.7566/JPSJ.86.054601} {\bibfield
  {journal} {\bibinfo  {journal} {J. Phys. Soc. Jpn.}\ }\textbf {\bibinfo
  {volume} {86}},\ \bibinfo {pages} {054601} (\bibinfo {year}
  {2017})}\BibitemShut {NoStop}%
\bibitem [{\citenamefont {Kontani}(2003)}]{PhysRevB.67.014408}%
  \BibitemOpen
  \bibfield  {author} {\bibinfo {author} {\bibfnamefont {H.}~\bibnamefont
  {Kontani}},\ }\href {https://doi.org/10.1103/PhysRevB.67.014408} {\bibfield
  {journal} {\bibinfo  {journal} {Phys. Rev. B}\ }\textbf {\bibinfo {volume}
  {67}},\ \bibinfo {pages} {014408} (\bibinfo {year} {2003})}\BibitemShut
  {NoStop}%
\bibitem [{\citenamefont {Kohno}\ \emph {et~al.}(2016)\citenamefont {Kohno},
  \citenamefont {Hiraoka}, \citenamefont {Hatami},\ and\ \citenamefont
  {Bauer}}]{PhysRevB.94.104417}%
  \BibitemOpen
  \bibfield  {author} {\bibinfo {author} {\bibfnamefont {H.}~\bibnamefont
  {Kohno}}, \bibinfo {author} {\bibfnamefont {Y.}~\bibnamefont {Hiraoka}},
  \bibinfo {author} {\bibfnamefont {M.}~\bibnamefont {Hatami}},\ and\ \bibinfo
  {author} {\bibfnamefont {G.~E.~W.}\ \bibnamefont {Bauer}},\ }\href
  {https://doi.org/10.1103/PhysRevB.94.104417} {\bibfield  {journal} {\bibinfo
  {journal} {Phys. Rev. B}\ }\textbf {\bibinfo {volume} {94}},\ \bibinfo
  {pages} {104417} (\bibinfo {year} {2016})}\BibitemShut {NoStop}%
\bibitem [{\citenamefont {Murakami}\ \emph {et~al.}(2003)\citenamefont
  {Murakami}, \citenamefont {Nagaosa},\ and\ \citenamefont
  {Zhang}}]{Murakami1348}%
  \BibitemOpen
  \bibfield  {author} {\bibinfo {author} {\bibfnamefont {S.}~\bibnamefont
  {Murakami}}, \bibinfo {author} {\bibfnamefont {N.}~\bibnamefont {Nagaosa}},\
  and\ \bibinfo {author} {\bibfnamefont {S.-C.}\ \bibnamefont {Zhang}},\ }\href
  {https://doi.org/10.1126/science.1087128} {\bibfield  {journal} {\bibinfo
  {journal} {Science}\ }\textbf {\bibinfo {volume} {301}},\ \bibinfo {pages}
  {1348} (\bibinfo {year} {2003})}\BibitemShut {NoStop}%
\bibitem [{\citenamefont {Murakami}\ \emph {et~al.}(2004)\citenamefont
  {Murakami}, \citenamefont {Nagaosa},\ and\ \citenamefont
  {Zhang}}]{PhysRevB.69.235206}%
  \BibitemOpen
  \bibfield  {author} {\bibinfo {author} {\bibfnamefont {S.}~\bibnamefont
  {Murakami}}, \bibinfo {author} {\bibfnamefont {N.}~\bibnamefont {Nagaosa}},\
  and\ \bibinfo {author} {\bibfnamefont {S.-C.}\ \bibnamefont {Zhang}},\ }\href
  {https://doi.org/10.1103/PhysRevB.69.235206} {\bibfield  {journal} {\bibinfo
  {journal} {Phys. Rev. B}\ }\textbf {\bibinfo {volume} {69}},\ \bibinfo
  {pages} {235206} (\bibinfo {year} {2004})}\BibitemShut {NoStop}%
\bibitem [{\citenamefont {Murakami}(2004)}]{PhysRevB.69.241202}%
  \BibitemOpen
  \bibfield  {author} {\bibinfo {author} {\bibfnamefont {S.}~\bibnamefont
  {Murakami}},\ }\href {https://doi.org/10.1103/PhysRevB.69.241202} {\bibfield
  {journal} {\bibinfo  {journal} {Phys. Rev. B}\ }\textbf {\bibinfo {volume}
  {69}},\ \bibinfo {pages} {241202(R)} (\bibinfo {year} {2004})}\BibitemShut
  {NoStop}%
\bibitem [{\citenamefont {Sinova}\ \emph {et~al.}(2004)\citenamefont {Sinova},
  \citenamefont {Culcer}, \citenamefont {Niu}, \citenamefont {Sinitsyn},
  \citenamefont {Jungwirth},\ and\ \citenamefont
  {MacDonald}}]{PhysRevLett.92.126603}%
  \BibitemOpen
  \bibfield  {author} {\bibinfo {author} {\bibfnamefont {J.}~\bibnamefont
  {Sinova}}, \bibinfo {author} {\bibfnamefont {D.}~\bibnamefont {Culcer}},
  \bibinfo {author} {\bibfnamefont {Q.}~\bibnamefont {Niu}}, \bibinfo {author}
  {\bibfnamefont {N.~A.}\ \bibnamefont {Sinitsyn}}, \bibinfo {author}
  {\bibfnamefont {T.}~\bibnamefont {Jungwirth}},\ and\ \bibinfo {author}
  {\bibfnamefont {A.~H.}\ \bibnamefont {MacDonald}},\ }\href
  {https://doi.org/10.1103/PhysRevLett.92.126603} {\bibfield  {journal}
  {\bibinfo  {journal} {Phys. Rev. Lett.}\ }\textbf {\bibinfo {volume} {92}},\
  \bibinfo {pages} {126603} (\bibinfo {year} {2004})}\BibitemShut {NoStop}%
\bibitem [{\citenamefont {Ando}\ \emph {et~al.}(2008)\citenamefont {Ando},
  \citenamefont {Takahashi}, \citenamefont {Harii}, \citenamefont {Sasage},
  \citenamefont {Ieda}, \citenamefont {Maekawa},\ and\ \citenamefont
  {Saitoh}}]{PhysRevLett.101.036601}%
  \BibitemOpen
  \bibfield  {author} {\bibinfo {author} {\bibfnamefont {K.}~\bibnamefont
  {Ando}}, \bibinfo {author} {\bibfnamefont {S.}~\bibnamefont {Takahashi}},
  \bibinfo {author} {\bibfnamefont {K.}~\bibnamefont {Harii}}, \bibinfo
  {author} {\bibfnamefont {K.}~\bibnamefont {Sasage}}, \bibinfo {author}
  {\bibfnamefont {J.}~\bibnamefont {Ieda}}, \bibinfo {author} {\bibfnamefont
  {S.}~\bibnamefont {Maekawa}},\ and\ \bibinfo {author} {\bibfnamefont
  {E.}~\bibnamefont {Saitoh}},\ }\href
  {https://doi.org/10.1103/PhysRevLett.101.036601} {\bibfield  {journal}
  {\bibinfo  {journal} {Phys. Rev. Lett.}\ }\textbf {\bibinfo {volume} {101}},\
  \bibinfo {pages} {036601} (\bibinfo {year} {2008})}\BibitemShut {NoStop}%
\bibitem [{\citenamefont {Liu}\ \emph {et~al.}(2011)\citenamefont {Liu},
  \citenamefont {Moriyama}, \citenamefont {Ralph},\ and\ \citenamefont
  {Buhrman}}]{PhysRevLett.106.036601}%
  \BibitemOpen
  \bibfield  {author} {\bibinfo {author} {\bibfnamefont {L.}~\bibnamefont
  {Liu}}, \bibinfo {author} {\bibfnamefont {T.}~\bibnamefont {Moriyama}},
  \bibinfo {author} {\bibfnamefont {D.~C.}\ \bibnamefont {Ralph}},\ and\
  \bibinfo {author} {\bibfnamefont {R.~A.}\ \bibnamefont {Buhrman}},\ }\href
  {https://doi.org/10.1103/PhysRevLett.106.036601} {\bibfield  {journal}
  {\bibinfo  {journal} {Phys. Rev. Lett.}\ }\textbf {\bibinfo {volume} {106}},\
  \bibinfo {pages} {036601} (\bibinfo {year} {2011})}\BibitemShut {NoStop}%
\bibitem [{\citenamefont {Shen}\ \emph {et~al.}(2014)\citenamefont {Shen},
  \citenamefont {Vignale},\ and\ \citenamefont
  {Raimondi}}]{PhysRevLett.112.096601}%
  \BibitemOpen
  \bibfield  {author} {\bibinfo {author} {\bibfnamefont {K.}~\bibnamefont
  {Shen}}, \bibinfo {author} {\bibfnamefont {G.}~\bibnamefont {Vignale}},\ and\
  \bibinfo {author} {\bibfnamefont {R.}~\bibnamefont {Raimondi}},\ }\href
  {https://doi.org/10.1103/PhysRevLett.112.096601} {\bibfield  {journal}
  {\bibinfo  {journal} {Phys. Rev. Lett.}\ }\textbf {\bibinfo {volume} {112}},\
  \bibinfo {pages} {096601} (\bibinfo {year} {2014})}\BibitemShut {NoStop}%
\bibitem [{\citenamefont {Tserkovnyak}\ \emph
  {et~al.}(2002{\natexlab{a}})\citenamefont {Tserkovnyak}, \citenamefont
  {Brataas},\ and\ \citenamefont {Bauer}}]{PhysRevLett.88.117601}%
  \BibitemOpen
  \bibfield  {author} {\bibinfo {author} {\bibfnamefont {Y.}~\bibnamefont
  {Tserkovnyak}}, \bibinfo {author} {\bibfnamefont {A.}~\bibnamefont
  {Brataas}},\ and\ \bibinfo {author} {\bibfnamefont {G.~E.~W.}\ \bibnamefont
  {Bauer}},\ }\href {https://doi.org/10.1103/PhysRevLett.88.117601} {\bibfield
  {journal} {\bibinfo  {journal} {Phys. Rev. Lett.}\ }\textbf {\bibinfo
  {volume} {88}},\ \bibinfo {pages} {117601} (\bibinfo {year}
  {2002}{\natexlab{a}})}\BibitemShut {NoStop}%
\bibitem [{\citenamefont {Tserkovnyak}\ \emph
  {et~al.}(2002{\natexlab{b}})\citenamefont {Tserkovnyak}, \citenamefont
  {Brataas},\ and\ \citenamefont {Bauer}}]{PhysRevB.66.224403}%
  \BibitemOpen
  \bibfield  {author} {\bibinfo {author} {\bibfnamefont {Y.}~\bibnamefont
  {Tserkovnyak}}, \bibinfo {author} {\bibfnamefont {A.}~\bibnamefont
  {Brataas}},\ and\ \bibinfo {author} {\bibfnamefont {G.~E.~W.}\ \bibnamefont
  {Bauer}},\ }\href {https://doi.org/10.1103/PhysRevB.66.224403} {\bibfield
  {journal} {\bibinfo  {journal} {Phys. Rev. B}\ }\textbf {\bibinfo {volume}
  {66}},\ \bibinfo {pages} {224403} (\bibinfo {year}
  {2002}{\natexlab{b}})}\BibitemShut {NoStop}%
\bibitem [{\citenamefont {Tserkovnyak}\ \emph {et~al.}(2005)\citenamefont
  {Tserkovnyak}, \citenamefont {Brataas}, \citenamefont {Bauer},\ and\
  \citenamefont {Halperin}}]{RevModPhys.77.1375}%
  \BibitemOpen
  \bibfield  {author} {\bibinfo {author} {\bibfnamefont {Y.}~\bibnamefont
  {Tserkovnyak}}, \bibinfo {author} {\bibfnamefont {A.}~\bibnamefont
  {Brataas}}, \bibinfo {author} {\bibfnamefont {G.~E.~W.}\ \bibnamefont
  {Bauer}},\ and\ \bibinfo {author} {\bibfnamefont {B.~I.}\ \bibnamefont
  {Halperin}},\ }\href {https://doi.org/10.1103/RevModPhys.77.1375} {\bibfield
  {journal} {\bibinfo  {journal} {Rev. Mod. Phys.}\ }\textbf {\bibinfo {volume}
  {77}},\ \bibinfo {pages} {1375} (\bibinfo {year} {2005})}\BibitemShut
  {NoStop}%
\bibitem [{\citenamefont {Ohe}\ \emph {et~al.}(2007)\citenamefont {Ohe},
  \citenamefont {Takeuchi},\ and\ \citenamefont
  {Tatara}}]{PhysRevLett.99.266603}%
  \BibitemOpen
  \bibfield  {author} {\bibinfo {author} {\bibfnamefont {J.-i.}\ \bibnamefont
  {Ohe}}, \bibinfo {author} {\bibfnamefont {A.}~\bibnamefont {Takeuchi}},\ and\
  \bibinfo {author} {\bibfnamefont {G.}~\bibnamefont {Tatara}},\ }\href
  {https://doi.org/10.1103/PhysRevLett.99.266603} {\bibfield  {journal}
  {\bibinfo  {journal} {Phys. Rev. Lett.}\ }\textbf {\bibinfo {volume} {99}},\
  \bibinfo {pages} {266603} (\bibinfo {year} {2007})}\BibitemShut {NoStop}%
\bibitem [{\citenamefont {Takeuchi}\ and\ \citenamefont
  {Tatara}(2008)}]{JPSJ.77.074701}%
  \BibitemOpen
  \bibfield  {author} {\bibinfo {author} {\bibfnamefont {A.}~\bibnamefont
  {Takeuchi}}\ and\ \bibinfo {author} {\bibfnamefont {G.}~\bibnamefont
  {Tatara}},\ }\href {https://doi.org/10.1143/JPSJ.77.074701} {\bibfield
  {journal} {\bibinfo  {journal} {J. Phys. Soc. Jpn.}\ }\textbf {\bibinfo
  {volume} {77}},\ \bibinfo {pages} {074701} (\bibinfo {year}
  {2008})}\BibitemShut {NoStop}%
\bibitem [{\citenamefont {Takeuchi}\ \emph {et~al.}(2010)\citenamefont
  {Takeuchi}, \citenamefont {Hosono},\ and\ \citenamefont
  {Tatara}}]{PhysRevB.81.144405}%
  \BibitemOpen
  \bibfield  {author} {\bibinfo {author} {\bibfnamefont {A.}~\bibnamefont
  {Takeuchi}}, \bibinfo {author} {\bibfnamefont {K.}~\bibnamefont {Hosono}},\
  and\ \bibinfo {author} {\bibfnamefont {G.}~\bibnamefont {Tatara}},\ }\href
  {https://doi.org/10.1103/PhysRevB.81.144405} {\bibfield  {journal} {\bibinfo
  {journal} {Phys. Rev. B}\ }\textbf {\bibinfo {volume} {81}},\ \bibinfo
  {pages} {144405} (\bibinfo {year} {2010})}\BibitemShut {NoStop}%
\bibitem [{\citenamefont {Hosono}\ \emph {et~al.}(2010)\citenamefont {Hosono},
  \citenamefont {Takeuchi},\ and\ \citenamefont {Tatara}}]{JPSJ.79.014708}%
  \BibitemOpen
  \bibfield  {author} {\bibinfo {author} {\bibfnamefont {K.}~\bibnamefont
  {Hosono}}, \bibinfo {author} {\bibfnamefont {A.}~\bibnamefont {Takeuchi}},\
  and\ \bibinfo {author} {\bibfnamefont {G.}~\bibnamefont {Tatara}},\ }\href
  {https://doi.org/10.1143/JPSJ.79.014708} {\bibfield  {journal} {\bibinfo
  {journal} {J. Phys. Soc. Jpn.}\ }\textbf {\bibinfo {volume} {79}},\ \bibinfo
  {pages} {014708} (\bibinfo {year} {2010})}\BibitemShut {NoStop}%
\bibitem [{\citenamefont {Kimata}\ \emph {et~al.}(2019)\citenamefont {Kimata},
  \citenamefont {Chen}, \citenamefont {Kondou}, \citenamefont {Sugimoto},
  \citenamefont {Muduli}, \citenamefont {Ikhlas}, \citenamefont {Omori},
  \citenamefont {Tomita}, \citenamefont {MacDonald}, \citenamefont
  {Nakatsuji},\ and\ \citenamefont {Otani}}]{Kimata2019}%
  \BibitemOpen
  \bibfield  {author} {\bibinfo {author} {\bibfnamefont {M.}~\bibnamefont
  {Kimata}}, \bibinfo {author} {\bibfnamefont {H.}~\bibnamefont {Chen}},
  \bibinfo {author} {\bibfnamefont {K.}~\bibnamefont {Kondou}}, \bibinfo
  {author} {\bibfnamefont {S.}~\bibnamefont {Sugimoto}}, \bibinfo {author}
  {\bibfnamefont {P.~K.}\ \bibnamefont {Muduli}}, \bibinfo {author}
  {\bibfnamefont {M.}~\bibnamefont {Ikhlas}}, \bibinfo {author} {\bibfnamefont
  {Y.}~\bibnamefont {Omori}}, \bibinfo {author} {\bibfnamefont
  {T.}~\bibnamefont {Tomita}}, \bibinfo {author} {\bibfnamefont {A.~H.}\
  \bibnamefont {MacDonald}}, \bibinfo {author} {\bibfnamefont {S.}~\bibnamefont
  {Nakatsuji}},\ and\ \bibinfo {author} {\bibfnamefont {Y.}~\bibnamefont
  {Otani}},\ }\href {https://doi.org/10.1038/s41586-018-0853-0} {\bibfield
  {journal} {\bibinfo  {journal} {Nature (London)}\ }\textbf {\bibinfo {volume}
  {565}},\ \bibinfo {pages} {627} (\bibinfo {year} {2019})}\BibitemShut
  {NoStop}%
\end{thebibliography}
\end{document}